\documentclass[%
 reprint,
superscriptaddress,
 prb,
]{revtex4-2}
\usepackage{graphicx}% Include figure files
\usepackage{dcolumn}% Align table columns on decimal point
\usepackage{bm}% bold math
\usepackage{float}
\usepackage{array}
\usepackage{booktabs}
\usepackage{amssymb}
\usepackage{bbding}
\usepackage[dvipsnames]{xcolor}

\usepackage[colorlinks=true,linkcolor=blue,anchorcolor=red,citecolor=blue,urlcolor=blue]{hyperref}

\usepackage[mathlines]{lineno}
\def\PH{\textcolor{black}}
\begin{document}
\title{Transverse Cooper-Pair Rectifier}
\author{Pei-Hao Fu}
\email{phy.phfu@gmail.com}
\affiliation{Science, Mathematics and Technology, Singapore University of Technology and Design, Singapore 487372, Singapore}
\affiliation{Department of Physics, National University of Singapore, Singapore 117542}

\author{Yong Xu}
%\email{xuyong_nbut@163.com}
\affiliation{Institute of Materials, Ningbo University of Technology, Ningbo 315016, China}

\author{Jun-Feng Liu}
\email{phjfliu@gzhu.edu.cn}
\affiliation{School of Physics and Materials Science, Guangzhou University, Guangzhou 510006, China}

\author{Ching Hua Lee}
\email{phylch@nus.edu.sg}
\affiliation{Department of Physics, National University of Singapore, Singapore 117542}

\author{Yee Sin Ang}
\email{yeesin\_ang@sutd.edu.sg}
\affiliation{Science, Mathematics and Technology, Singapore University of Technology and Design, Singapore 487372, Singapore}

%%%%%%%%%%%%%%%%%%%%%%%%%%%%%%%%%%%%%%%%%%%%%%%%%%%%%%%%%%%%%%%%%%%%%%
\begin{abstract}
Non-reciprocal devices are key components in modern electronics covering broad applications ranging from transistors to logic circuits thanks to the output rectified signal in the direction parallel to the input.
In this work, we propose a transverse Cooper-pair rectifier in which a non-reciprocal current is perpendicular to the driving field, when inversion, time reversal, and mirror symmetries are broken simultaneously. 
The Blonder-Tinkham-Klapwijk formalism is developed to describe the transverse current-voltage relation in a normal-metal/superconductor tunneling junction, where symmetry constraints are achieved by an effective built-in supercurrent manifesting in an asymmetric and anisotropic Andreev reflection.
The asymmetry in the Andreev reflection is induced when inversion and time reversal symmetry are broken by the supercurrent component parallel to the junction while the anisotropy occurs when the mirror symmetry with respect to the normal of the junction interface is broken by the perpendicular supercurrent component to the junction.
Compared to the conventional longitudinal one, the transverse rectifier supports fully polarized diode efficiency and colossal nonreciprocal conductance rectification, completely decoupling the path of the input excitation from the output rectified signal. 
This work provides a formalism for realizing transverse non-reciprocity in superconducting junctions, which is expected to be achieved by modifying current experimental setups and may pave the way for future low-dissipation superconducting electronics.
\end{abstract}
\maketitle
\begin{figure}[t]
\centering \includegraphics[width=0.48\textwidth]{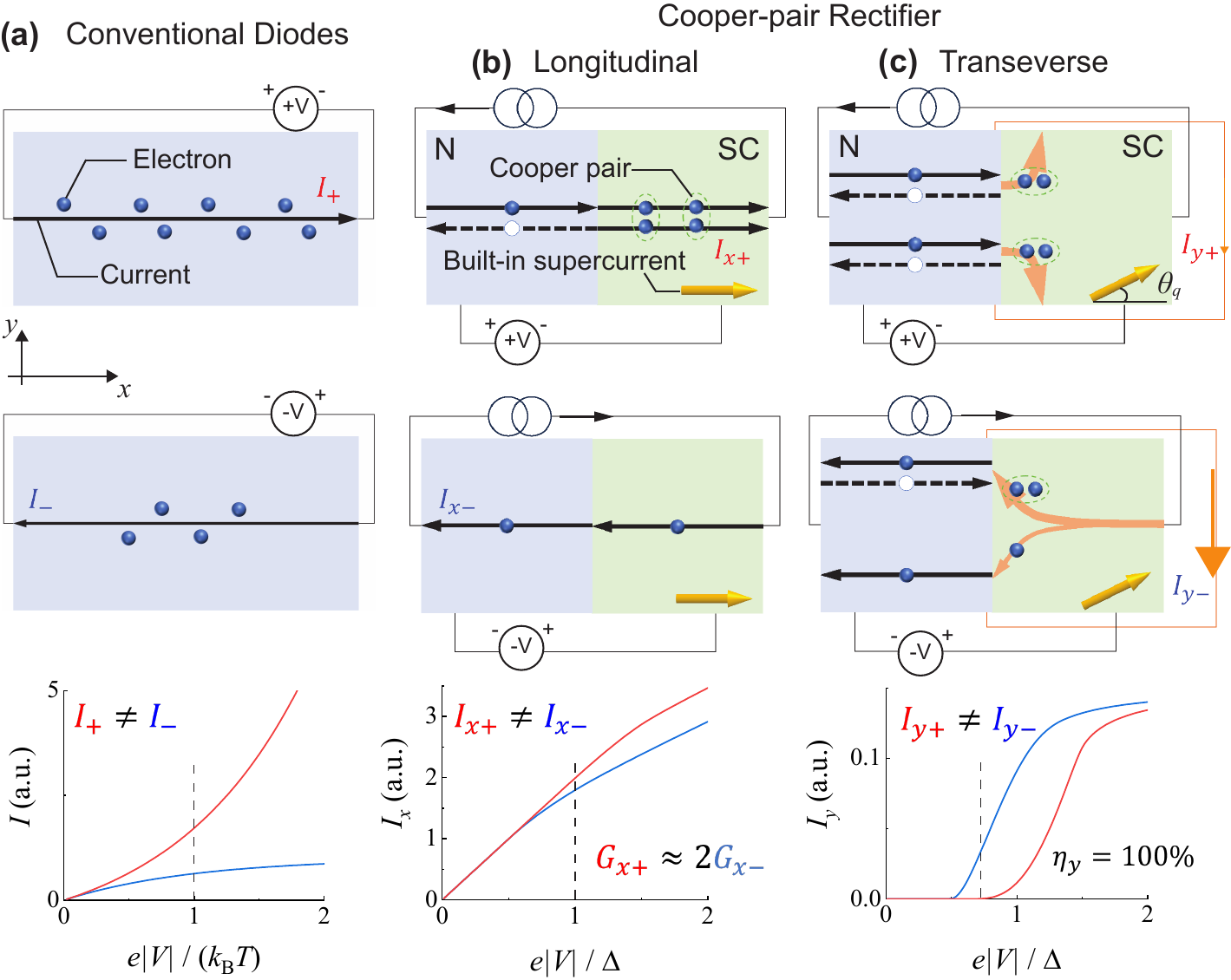}
\caption{
Schematic transport and $I$-$V$ relations of the conventional electronic diode, $I=\exp(eV/k_BT)-1$ \cite{Sze2012} and two types of Cooper-pair rectifier [Eq. (\ref{eq_iv})] assisted by the built-in supercurrent flowing deflected from the junction direction ($+x$) by an angle $\theta_q$, \PH{with a voltage drop $V$ due to current bias.}
Negative-biased currents are flipped to the positive side for comparison in an arbitrary unit (a.u.).
The doubling conductance $G_{x+}=2G_{x-}$ indicates the direction-selective Cooper-pair (paired electrons) transferring process. 
Fully polarized rectified current efficiency [Eq. (\ref{eq_eta})] is expected in the transverse Cooper-pair rectifier.
}
\label{fig_diode}
\end{figure}

%%%%%%%%%%%%%%%%%%%%%%%%%%%%%%%%%%%%%%%%%%%%%%%%%%%%%%%%%%%%%%%%%%%%%%%%%%%%%
\textit{\textcolor{blue}{Introduction.-}} 
As fundamental electronic components, diodes conduct current primarily in one direction, offering low (high) resistance in the forward (reverse) bias \cite{Sze2012,Ideue2017}. 
The rectified current direction is usually \textit{parallel} to the applied bias, which is dubbed as the longitudinal diode effect [Fig. \ref{fig_diode}(a)].
The longitudinal diodes play essential roles from consumer electronics and signal processors \cite{Sze2012} to quantum sensing and computing \cite{Shukla2023quantum}.
New device architectures based on molecules \cite{chen2017molecular}, superconductors [Fig. \ref{fig_diode}(b)] \cite{Nadeem2023,Wakatsuki2017}, and other quantum materials \cite{Nagaosa2024,Tokura2018} are being actively explored to achieve giant rectification \cite{Arenas2024,lee2020giant,pereira2019perfect} at low-energy consumption \cite{zhao2023engineering}.
%%%%%%%%%%%%%%%%%%%%
%%%%% Figure 1 %%%%%
%%%%%%%%%%%%%%%%%%%%

Beyond the longitudinal diodes, the recently revealed Hall rectifiers \cite{Isobe2020,Min2023,Onishi2022,Daido2023} support output DC signals \textit{perpendicular} to the input AC ones, leveraging the nonlinear Hall response of noncentrosymmetric quantum materials \cite{Onishi2022,Daido2023,Qin2024}. 
The decoupling between output and input direction enables a colossal Hall rectification effect, which can convert terahertz electromagnetic waves into DC output with high rectification efficiency and low-energy consumption \cite{Min2023}.

In this work, we implement the concept of a transverse rectifier to the superconducting regime and propose a \textit{transverse Cooper-pair rectifier} [Fig. \ref{fig_diode}(c)], where  the Cooper-pair-dominated current is perpendicular to bias with a direction-selective magnitude. 
This transport effect is distinct from the longitudinal superconducting rectification \cite{Nadeem2023,Wakatsuki2017}, and based on a mechanism different from single-electron nonlinear Hall rectifiers \cite{Isobe2020,Min2023,Onishi2022,Daido2023}. 
Our proposed setup is a normal metal-superconductor (N-S) junction and, thus does not intrinsically require noncentrosymmetric quantum materials \cite{Isobe2020,Min2023,Onishi2022} or fluctuating Cooper pairs \cite{Daido2023} as components.
As a superconducting tunneling junction \cite{Strambini2022,Araujo2023,Geng2023}, our proposed rectifier is a low-dispassion cryogenic electronics device in low working bias ($\sim $meV) and temperature ($\sim $mK) conditions \cite{Braginski2019} with flexible external electromagnet knots and configurations.
The device performance is featured by unidirectional Cooper-pair current and giant conductance rectification.

\textit{\textcolor{blue}{Symmetry constraints.-}} 
To characterize the transverse Cooper-pair rectifier, our work aims to develop a universal longitudinal and transverse current-voltage ($I_{x,y}$-$V$) relation, which can be conceptually explained through the symmetry constraints (Table \ref{Table_Sym}).  
Generally, the linearized $I_{x,y}$-$V$ relation of a two-dimensional tunneling junction with a bias $V$ along $x$-direction is \cite{Jacquod2012}
\begin{equation}
I_{x,y}(eV) = G_{x,y}( eV) V\text{,}
\label{Eq_liniv}
\end{equation}%
where $G_{x,y}( eV) =G_{0}\int_{-\pi /2}^{\pi /2}d\theta_{k}\cos \theta _{k}\hat{\bm{k}}_{F_{x,y}}T_{x,y}( eV,\theta _{k})$
is the longitudinal ($x$) and transverse ($y$) conductance in a unit of $G_{0}$, $\bm{\hat{k}}_{F}=\bm{k}_{F}/k_{F}=( \cos \theta _{k},\sin \theta _{k}) $ denotes the direction of the incident electrons measured from $x$-axis with $|\theta _{k}| \leq \pi/2$, and $T_{x/y}(eV,\theta _{k}) $ is the direction-resolved transmission probability parallel/perpendicular to the bias. 
\PH{In semiconducting \cite{Sze2012} and dissipative superconducting regime \cite{Hu2007,Misaki2021},} the breaking of inversion symmetry $\mathcal{I}_{x}$ causes a longitudinal non-reciprocal current with $I_{x}(eV) \neq I_{x}( -eV)$ due to the \textit{asymmetric} conductance $G_{x}( eV) \neq G_{x}( -eV) $ with respect to zero bias.
%in semiconducting due to the \textit{asymmetric} conductance $G_{x}( eV) \neq G_{x}( -eV) $ with respect to zero bias \cite{Sze2012} and \PH{dissipative superconducting regime \cite{Hu2007,Misaki2021}.}
%The further breaking of time-reversal symmetry $\mathcal{T}$ induces the longitudinal superconducting rectification \PH{in dissipationless regime }\cite{Misaki2021,Zinkl2022,Hu2007,Wang2022} embodying in both $I_{x}( eV) \neq I_{x}( -eV) $ and $G_{x}( eV) =2G_{x}( -eV) $. 
%\PH{The factor $2$ denoting unidirectional collective drifting motion of the Cooper pairs breaking $\mathcal{T}$ in BCS theory \cite{Bagwell1994}.}

\PH{The further breaking of the mirror symmetry to the bias direction $\mathcal{M}_{y}$ leads to the transverse rectification.}
The breaking of $\mathcal{M}_{y}$ generates a transverse current, by destroying the initial equivalent contribution from the upward-going ($+\theta _{k}$) and downward-going ($-\theta _{k}$) modes [Fig. \ref{fig_diode}]. 
A non-zero $I_y$ is found due to the \textit{anisotropic} transmission probability $T_{y}( eV,\theta _{k}) \neq T_{y}(eV,-\theta _{k}) $.
This universal argument is valid for all tunneling Hall (transverse) \PH{currents} in both non-superconducting \cite{Scharf2016} and superconducting junction \cite{Ren2014,Salehi2023,Zhou1998,Parafilo2023}.
Accompanied by the \PH{breaking $\mathcal{I}_x$}, the anisotropic and asymmetric transmission probability leads to the transverse rectification with $I_{y}( eV) \neq I_{y}( -eV) $.

%The transverse Cooper-pair rectification effect occurs when the mirror symmetry to the bias direction $\mathcal{M}_{y}$ is broken simultaneously with $\mathcal{I}_{x}$ and $\mathcal{T}$. 
%The breaking of $\mathcal{M}_{y}$ generates a transverse current, by destroying the initial equivalent contribution from between upward-going ($+\theta _{k}$) and downward-going ($-\theta _{k}$) modes [Fig. \ref{fig_diode}]. 
%A non-zero $I_y$ is expected due to the anisotropic transmission probability $T_{y}( eV,\theta _{k}) \neq T_{y}(eV,-\theta _{k}) $.
%This universal argument is valid for all tunneling Hall (transverse) \PH{currents} in both non-superconducting \cite{Scharf2016} and superconducting junction \cite{Ren2014,Salehi2023,Zhou1998,Parafilo2023}.
%Accompanied by the breaking $\mathcal{I}_x$ and $\mathcal{T}$, the anisotropic and asymmetric transmission probability leads to the transverse Cooper-pair rectification effect with $I_{y}( eV) \neq I_{y}( -eV) $.

The universal symmetry constraints above are specifically manifested in the Cooper-pair rectifier in an N-S junction.
By developing the transverse current within the Blonder-Tinkham-Klapwijk (BTK) formalism \cite{Blonder1982}, the transmission probabilities of the N-S junction is \cite{sm}
\begin{equation}
T_{x(y)}(E,\theta _{k})=1-(+)R(E,\theta _{k})+R_{a}(E,\theta _{k})\text{,}
\label{eq_T}
\end{equation}%
which contains the effect of (i) the retro-Andreev (electron-hole) reflection $R_{a}(E,\theta _{k})$ always contributing to current by forming Cooper pairs and (ii) the specular normal (electron-electron) reflection $R(E,\theta _{k})$ which reduces the longitudinal current because of the backscattering towards the N lead, but enhances the transverse one near the N-S interface.

The symmetry constraints can be achieved by an effective built-in unidirectional supercurrent \cite{Bagwell1994,Canizares1995,Canizares1997,Riedel1999,Lukic2007}, whose component parallel to biased direction break $\mathcal{I}_{x}$ \cite{,Zinkl2022,Wang2022} and the perpendicular one breaks $\mathcal{M}_{y}$.
\PH{The time-reversal symmetry $\mathcal{T}$ is also broken because of the unidirectional collective drifting motion \cite{Bagwell1994}.}
The effective built-in unidirectional supercurrent is essential for superconducting non-reciprocity in both non-centrosymmetric superconductors \cite{Zhang2020,Ando2020,Bauriedl2022,Itahashi2020,Yasuda2019,JMasuko2022,Hou2022,Wakamura2024,Narita2022,Hoshino2018,Wakatsuki2018,Scammell2022,Ili2022,Daido2022a,Legg2022,Zaccone2024Theory} and Josephson junctions
\cite{Davydova2022,Valentini2024,Fominov2022,Souto2022,Fukaya2024,Ciaccia2023,Costa2022,Costa2023,Mazur2022,Turini2022,Margineda2023a,Reinhardt2023,Jeon2022,Baumgartner2022a,Baumgartner2022b,Lu2022,Tanaka2022,Cayao2023,Lotfizadeh2023,Alidoust2021,Amundsen2022a,Fu2022b,Pal2022,Trahms2022,Margineda2023b,Wu2022,Wei2022,Xie2022,Zhang2022,Hu2022,Legg2023,Wang2024,Zazunov2023,Zazunov2024,Ding2022,Ding2023,Bozkurt2023Double,Yerin2024Supercurrent,Banerjee2024Altermagnetic,Yu2024Time}, whose origination is diverse including finite-momentum Cooper pairs \cite{Yuan2018,Yuan2021,Hart2017} induced by Meissner effect \cite{Davydova2022}, quantum interference effect \cite{Valentini2024,Ciaccia2023,Fominov2022,Souto2022}, vortex currents \cite{Fukaya2024}, the interplay between spin-orbit coupling and in-plane Zeeman field \cite{Costa2022,Costa2023,Mazur2022,Turini2022,Margineda2023a,Reinhardt2023,Baumgartner2022a,Baumgartner2022b,Jeon2022,Lu2022,Tanaka2022,Cayao2023,Lotfizadeh2023,Alidoust2021,Amundsen2022a}, and the non-zero net velocity between paired electrons in asymmetric helical states \cite{Fu2022b}. 

%%%%%%%%%%%%%%%%%%%%%%%%%%%%%%%%%%%
%%%%%%% Symmetry table %%%%%%%
%%%%%%%%%%%%%%%%%%%%%%%%%%%%%%%%%%%
\begin{table}[t]
\caption{The symmetry constraints for difference rectifiers, including inversion symmetry $\mathcal{I}_{x}$, and mirror symmetry $\mathcal{M}_y$. Symbols \textcolor{red}{\XSolidBrush} and \textcolor{Green}{\Checkmark} denote symmetry breaking and preservation, respectively. 
\PH{Time-reversal symmetry $\mathcal{T}$ is further broken in the Cooper-pair rectifier in the current work due to the effective supercurrent.}}
\label{Table_Sym}
\setlength{\tabcolsep}{3mm}{
\begin{tabular}{c| c  cc} 
\toprule
 \textbf{Device} 
& \multicolumn{1}{c}{Conventional} & \multicolumn{2}{c}{Cooper-pair rectifier} \\ 
& diode 
& longitudinal  & transverse \\ 
\midrule
$\mathcal{I}_{x}$ & \textcolor{red}{\XSolidBrush} & \textcolor{red}{\XSolidBrush} & \textcolor{red}{\XSolidBrush} 
\\ 
$\mathcal{M}_y$ & \textcolor{Green}{\Checkmark} & \textcolor{Green}{\Checkmark} & \textcolor{red}{\XSolidBrush}
%\\ 
%$\mathcal{T}$ & \textcolor{Green}{\Checkmark} & \textcolor{red}{\XSolidBrush} & \textcolor{red}{\XSolidBrush} 
\\
\bottomrule
\end{tabular}}
\end{table}

\begin{figure}[t]
\centering \includegraphics[width=0.48\textwidth]{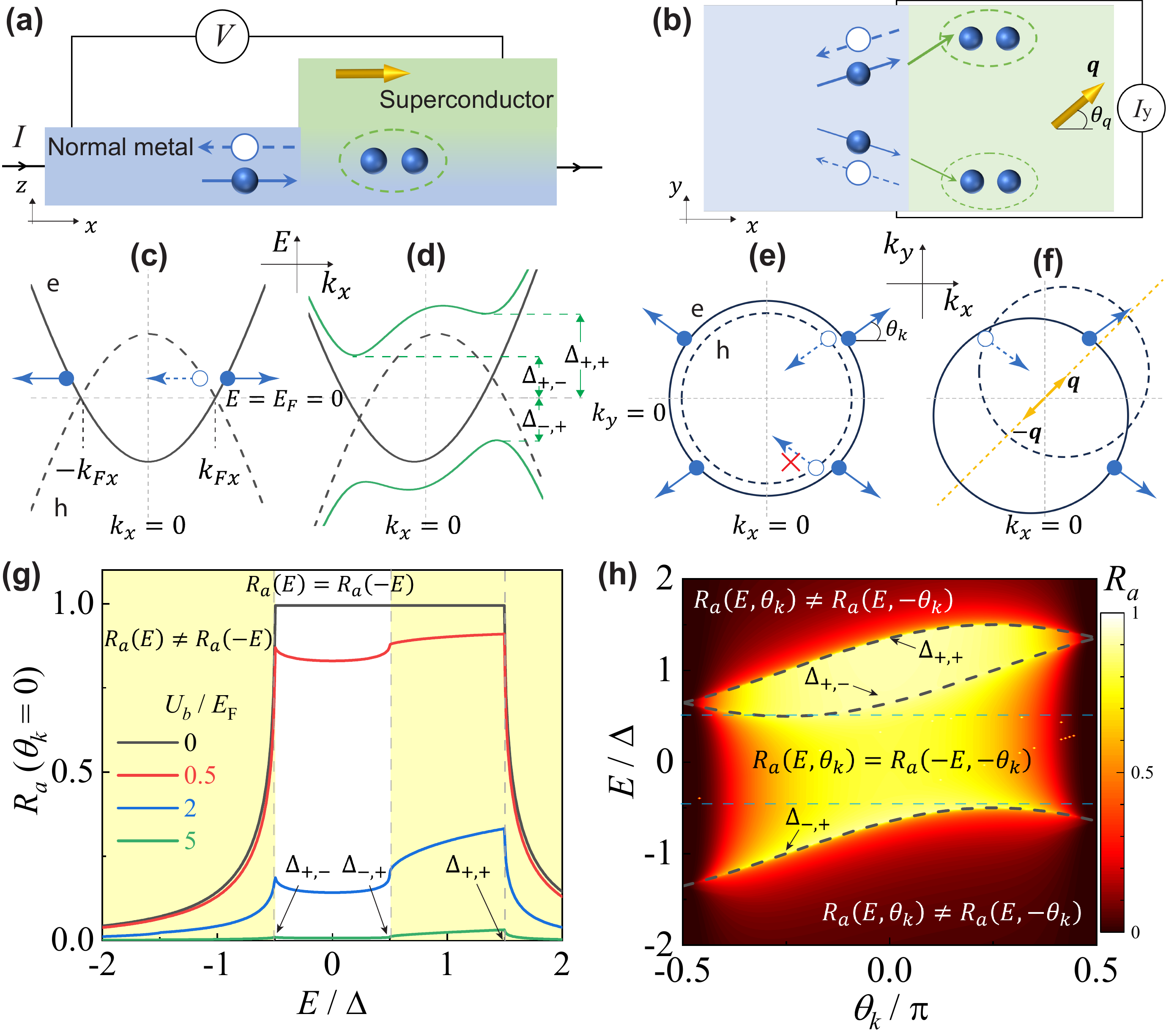}
\caption{Mechanism of the Cooper-pair rectifier explained by the (c, d) asymmetric and (e, f) anisotropic BdG spectrum and (g, h) the resulting AR spectroscopy.
The schematic (a) side and (b) top views of the biased N-S junction with a built-in supercurrent with a drifting Cooper-pair momentum $\bm{q}$. 
The solid (hollow) blue circles are electrons (holes) and the Cooper pairs composited by two correlated electrons are encircled by the green dashed lines. 
The blue arrows represent the motions of the quasi-particles. 
(c, d) Schematic dispersion [Eq. (\ref{eq_Edis})] and (e, f) Fermi surface of N and S affected by the built-in supercurrent. 
The solid (dashed) lines denote the electron (hole) bands and the green lines denote the BdG dispersion with multiple band edges $\Delta_{\pm,\pm}(\theta_{k})$ [Eq. (\ref{eq_delta})]. 
(g) The AR spectroscopy [Eq. (\ref{eq_ra})] of a normally incident electron ($\theta_k=0$) with different tunneling barriers. 
(h) The energy and incident-direction resolved AR probability with $V=0.5E_F$. 
Hereafter, we choose the Fermi energy \PH{$E_F=1$} as the energy unit and set $\hbar^2/(2m)=1$ keeping the Fermi wave vector $k_F=1$. 
The superconducting order parameter is \PH{$\Delta/E_F = 10^{-3}  $}. 
The magnitude of the Cooper-pair momentum is $q=0.5q_c$ and the direction is (g) $\theta_q=0$ and (h) $\theta_q=\pi/4$.
}
\label{fig_mechanism}
\end{figure}

%%%%%%%%%%%%%%%%%%%%%%%%%%%%%%%%%%%%%%%%%%%%%%%%%%%%%%%%%%%%%%%%%%%%%%%%%%%%%
\textit{\textcolor{blue}{BdG spectrum with built-in supercurrent.-}}
In a microscopic picture, the effective built-in unidirectional supercurrent modulates the Bogoliubov-de Gennes (BdG) spectrum, leading to an anisotropic and asymmetric transmission probability. 
The mechanism is exhibited in Fig. \ref{fig_mechanism}. 

To elaborate on the relation between built-in unidirectional supercurrent and the transmission probability, we first introduce this asymmetric and anisotropic BdG spectrum in an \textit{isotropic} $s$-wave superconductor. 
The Hamiltonian after a gauge transformation in Nambu space is \cite{Bagwell1994,Gennes1966} 
\begin{equation}
H_{\bm{q}}( \bm{k}) =\left( 
\begin{array}{cc}
h_{\text{e}}( \bm{k})  & \Delta  \\ 
\Delta & h_{\text{h}}( \bm{k}) 
\end{array}%
\right) \text{,}  \label{eq_hq}
\end{equation}%
where $\Delta$ is the superconducting gap coupling the electron-like $h_{\text{e}}( \bm{k}) =\hbar ^{2}/(2m)| \bm{q+k}| ^{2}-E_{F}$ and $h_{\text{h}}( \bm{k}) =-h_{\text{e}}^{\ast }( -\bm{k}) $ and hole-like $h_{\text{h}}( \bm{k}) =-h_{\text{e}}^{\ast }( - \bm{k}) $ quasiparticle with a Fermi level $E_{F}$, momentum $\bm{k}=(k_{x},k_{y})$, and Cooper-pair drifting momentum $\bm{q}=(q_{x},q_{y})=q( \cos \theta _{q},\sin \theta_{q}) $ \cite{Bagwell1994}. 

The spectrum of the BdG quasiparticle,
\begin{equation}
E_{\pm }^{\bm{q}}( \bm{k}) =\PH{\hbar ^{2}/m( \bm{%
k\cdot q})} \pm \sqrt{\epsilon _{q}^{2}+\Delta ^{2}}\text{,}
\label{eq_Edis}
\end{equation}%
\PH{with $\epsilon _{q}=\hbar ^{2}/(2m)(k^2+q^2)-E_F$} becomes asymmetric [Fig. \ref{fig_mechanism}(d)] and anisotropic [Fig. \ref{fig_mechanism}(f)], as the built-in supercurrent shifts dispersion centers of electrons (holes) from the origin [Fig. \ref{fig_mechanism}(c, e)] to $-\bm{q}$ ($+\bm{q}$) [Fig. \ref{fig_mechanism}(d, f)].
An anisotropy occurs because the original isotropic Fermi surfaces reduce to mirror-symmetric to the direction of the supercurrent. 
Subsequently, the symmetry between the electron and hole dispersions to $E=0$ is destroyed, resulting in an asymmetric BdG spectrum with an indirect gap.

%%%%%%%%%%%%%%%%%%%%
%%%%% Figure 2 %%%%%
%%%%%%%%%%%%%%%%%%%%

As exhibited in Fig. \ref{fig_mechanism}(d) and (h), the asymmetric and asymmetric BdG spectrum is characterized by four $\theta _{k}$-dependent band edges
\begin{equation}
\Delta _{\pm ,\pm }( \theta _{k}) =\pm \Delta +D_{\pm }(
\theta _{k}) \text{,}  \label{eq_delta}
\end{equation}
where the direction-selective Doppler energy shift \cite{Tkachov2015,Scharf2021,Dolcini2015,Fu2022a,Fu2022b,Rohlfing2009}
\begin{equation}
D_{\pm }( \theta _{k}) =\frac{\hbar ^{2}}{2m}[ q^{2}\pm
2qk_{F}\cos ( \theta _{q}\mp \theta _{k}) ] \text{,}
\label{eq_dE}
\end{equation}%
enhances/reduces the energy of states propagating parallel/anti-parallel to the built-in supercurrent. 
Thereby, the center of the band gap gains a $\theta _{k}$-dependent shift from $E=0$ to $D_{\pm}( \theta _{k}) $ at $\pm k_{Fx}=\pm k_{F}\cos \theta _{k}$, resulting in an asymmetric and asymmetric BdG spectrum.
Notably, BdG spectrum is gapless as $q\geq q_{c}=k_F\Delta/(2E_F)$ with $q_{c}$ \PH{as the depairing momentum at zero-temperature \cite{Bagwell1994,Riedel1999}}. 
The effective built-in supercurrents in recent experiments are inherited from the proximity effect \cite{Davydova2022,Zhu2021}, thereby do not have to obey the self-consistency equation \cite{Canizares1995,Canizares1997,Riedel1999,Bagwell1994,Lukic2007}.

%%%%%%%%%%%%%%%%%%%%%%%%%%%%%%%%%%%%%
%%%%%%%%%%%%%%%%%%%%%%%%%%%%%%%%%%%%%
%%%%%%%%%%%%%%%%%%%%%%%%%%%%%%%%%%%%%

\textit{\textcolor{blue}{Asymmetric and anisotropic Andreev reflection.-}}
The configuration of the BdG spectrum implies that the Andreev reflection (AR) probability is asymmetric and anisotropic and eventually causes a non-reciprocal current in the N-S junction with a prioritized transport direction provided by the built-in supercurrent.
By solving the quantum tunneling problem in the N-S junction, the AR probability for incident electrons with energy $E$ and direction $\theta _{k}$ is 
\cite{sm}
\begin{equation}
R_{a}(E,\theta _{k})=| 4\gamma _{-}^{\text{e}}\gamma
_{+}^{\text{h}}k_{Fx}^{2}Z^{-1}| ^{2}\text{,}  \label{eq_ra}
\end{equation}%
where $Z=\gamma _{+}^{\text{e}}\gamma _{+}^{\text{h}}[u_{b}^{2}+(2k_{Fx}+q_{x})^{2}]-\gamma _{-}^{\text{e}}\gamma _{-}^{\text{h}}(u_{b}^{2}+q_{x}^{2})$, $\gamma _{\tau =\pm 1}^{\text{e(h)}}=\sqrt{1+(-)\tau \Omega _{\tau }/\epsilon _{\tau }}$, $\Omega _{\tau}=sign(E-D_{\tau })\sqrt{\epsilon _{\tau }^{2}-\Delta ^{2}}$, $u_{b}=2mU_{b}/\hbar ^{2}$ and $\epsilon _{\tau }=E-D_{\tau }-\hbar^{2}q^{2}/(2m)$. 
To obtain Eq. (\ref{eq_ra}), the following assumptions are implemented. 
(i) Translational symmetry in $y$ direction is preserved. 
(ii) The effective mass $m$ and the Fermi level $E_{F}$ are shared by both the N and S sides. 
(iii) The interface tunnel barrier $U_{b}$ is the only boundary condition involved. 
(iv) Andreev approximation \cite{Andreev1964,Blonder1982} is considered, i.e. $E_{F}\gg $ $\Delta $, $\hbar ^{2}q^{2}/(2m)$. 
Thereby, the Fermi wave number mismatch between the N and S region is negligible, and the effect of the built-in supercurrent is revealed in the Dopper energy shift $D_{\pm}$. 
Approximation (iv) is valid because quasiparticles with momentum nearly parallel to the interface ($\theta _{k}\sim \pi /2$) do not contribute significantly to both the longitudinal and transverse current. 
Thereby their effects in the $I$-$V$ relation and related quantities will thus be negligible \cite{Mortensen1999}. 
Our results are checked numerically \cite{Fu2020} beyond the assumption (i-iv) and are qualitatively consistent with the analytical ones \cite{sm}.

The asymmetric and anisotropic AR characterizing the direction-selective superconducting gap in the BdG spectrum are exhibited in Fig. \ref{fig_mechanism}(g) and (h).
Asymmetry activated by $q_x$ manifests in $R_{a}(E,\theta _{k})\neq R_{a}(-E,\theta _{k})$ even with finite interfacial barriers $U_{b}$ [Fig. \ref{fig_mechanism}(g)], since $\mathcal{I}_x$ and $\mathcal{T}$ are broken. 
AR dominates in the asymmetric energy window $E\in \lbrack \Delta _{-,+}(\theta_{k}),\Delta _{+,+}(\theta_{k}) ]$ with Cooper-pair transferring, while exponentially decreases beyond this regime with increasing single-electron transmission.
Furthermore, anisotropic embodying in $R_{a}(E,\theta _{k})\neq R_{a}(E,-\theta _{k})$ is induced by $q_y$, which breaks $\mathcal{M}_y$ by providing a prioritized supercurrent orientation. 
Electrons with an incident direction closer to the supercurrent will confront a higher Doppler energy shift. 

Remarkably, the symmetry constraints for the transverse Cooper-pair rectifier are met when the supercurrent deviates from the direction of the junction ($q_{x,y}\neq0$). 
The simultaneous breaking of $\mathcal{I}_x$, $\mathcal{T}$, and $\mathcal{M}_y$ manifesting in asymmetric and anisotropic AR, i.e. $R_{a}(E,\theta _{k})\neq R_{a}(-E,\theta _{k})$ and $R_{a}(E,\theta _{k})\neq R_{a}(E,-\theta _{k})$. 
As a result, the transverse current occurs due to the imbalance AR between incident electrons with $\theta _{k}>0$ and $\theta _{k}<0$, and of distinctive responses to the bias direction.

%%%%%%%%%%%%%%%%%%%%
%%%%% Figure 3 %%%%%
%%%%%%%%%%%%%%%%%%%%

\begin{figure}[t]
    \centering \includegraphics[width=0.48\textwidth]{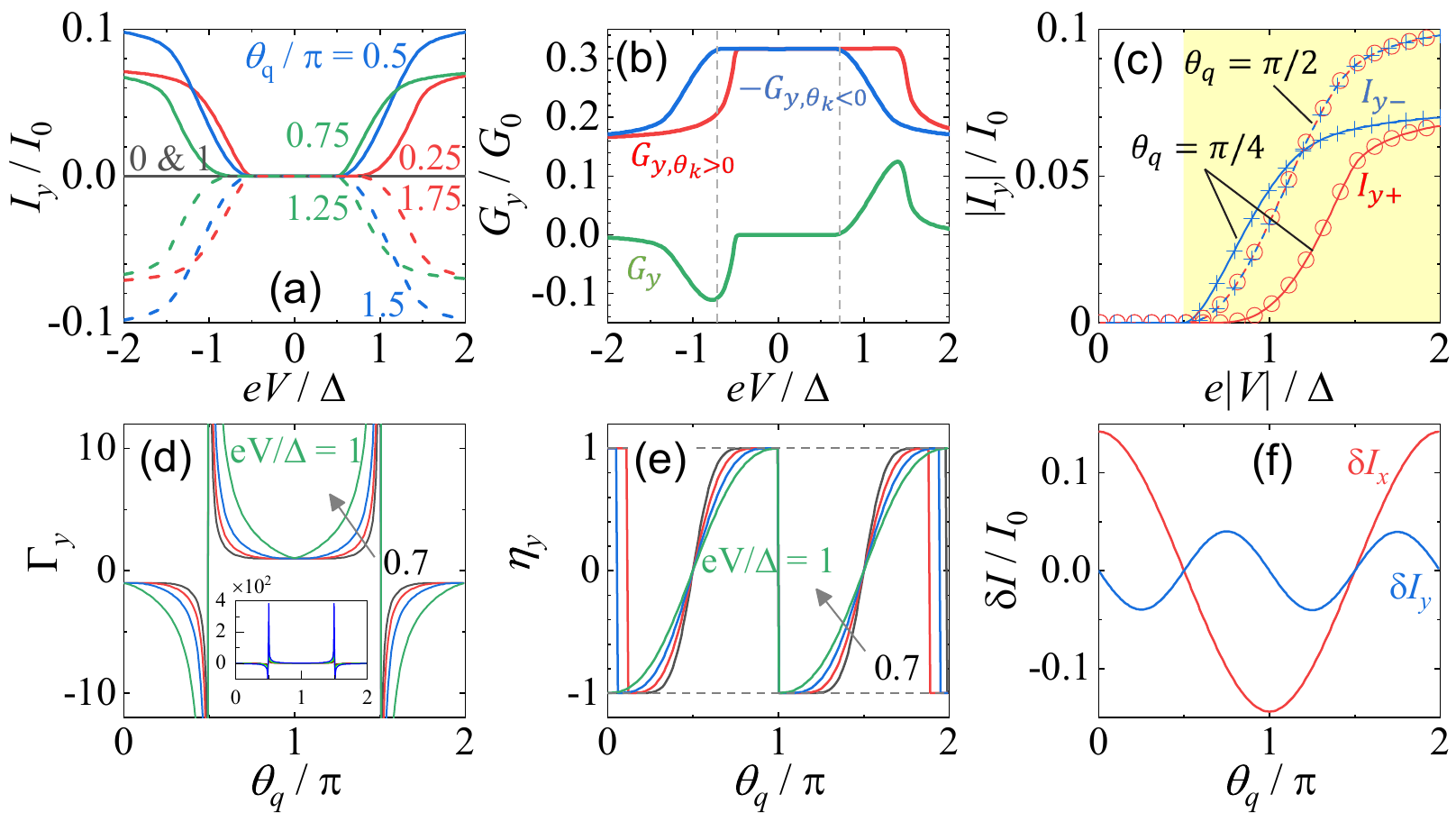}
    \caption{Non-reciprocal transport performance of transverse Cooper-pair rectifier. 
    Top panel: (a) $I_y$-$V$ relations [Eq. (\ref{eq_iv})] for various built-in supercurrent directions ($\theta_q$) and (b) differential conductance $G_{y}$ \PH{with $\theta_q=\pi/4$}, in the unit of $I_{0}=eWk_{F}v_{F}\Delta /(2\pi)$ and $G_{0}=eI_{0}$. (c) The negative-bias currents are flipped to compare with the positive-bias ones.  
    Bottom panel: The performance of the Cooper-pair diode to the built-in supercurrent direction ($\theta_q$) is revealed in (d) The \PH{conductance} difference [Eq. (\ref{eq_dixy})] (e) current efficiency [Eq. (\ref{eq_eta})] and (f) \PH{current} efficiency [Eq. (\ref{eq_gg})].
    Fully polarized transverse diode current ($\eta_y=100\%$) and a colossal nonreciprocal conductance rectification [inserted in (d)] ($|\Gamma_y|\rightarrow\infty$) are expected. 
    In (d), the bias is $e|V|/\Delta=1$. 
    The temperature is $k_BT=0.01\Delta$. 
    Other parameters are the same as those in Fig. \ref{fig_mechanism}.}
    \label{fig_current}
\end{figure}

%%%%%%%%%%%%%%%%%%%%%%%%%%%%%%%%%%%%%
%%%%%%%%%%%%%%%%%%%%%%%%%%%%%%%%%%%%%
%%%%%%%%%%%%%%%%%%%%%%%%%%%%%%%%%%%%%
\textit{\textcolor{blue}{Non-reciprocal $I$-$V$ relation and efficiency.-}}
To investigate the currents in a
biased N-S junction, we develop the non-reciprocal $I$-$V$ relation by following the quasiparticle current argument in the BTK formalism  \cite{Blonder1982,Datta1996}. 
By considering electron and hole currents and the quasiparticle probability conservation, the $I_{\beta=x,y}$-$V$ relation is \cite{sm}
\begin{eqnarray}
I_{\beta}( eV)  &=&\frac{1}{e}\int_{-\infty
}^{+\infty }dE G_{\beta}(E) [f_{\text{N}}( E,eV)-f_{\text{S}}( E,eV)]  \text{,} \label{eq_iv}
\end{eqnarray}%
where $G_{0}=eI_{0}$, $I_{0}=eW\hbar k_{F}^{2}/(\pi m)$ and $W$ is the width of the sample, and $f_{\text{S}}( E,eV) =f_{0}( E) =[1+\exp (E/k_{B}T)]^{-1}$ and $f_{\text{N}}( E) =f_{0}( E-eV) $ are the Fermi-Dirac distribution between S and N sides, respectively. 
Eq. (\ref{Eq_liniv}) is restored by linearized Eq. (\ref{eq_iv}) in small bias and low-temperature conditions \cite{sm}. 
The transverse current $I_{y}$ occurs when $\mathcal{M}_y$ is broken by $q_y$ [in Fig. \ref{fig_current}(a)] due to the net contribution between anisotropic AR as the bias exceeds the regime $[\max | \Delta _{-,+}( \theta _{k}) |,\min| \Delta _{+,-}( \theta _{k}) | ] $ [Fig. \ref{fig_mechanism}(g)].

A transverse Cooper-pair rectifier is found when $\mathcal{T}$ and $\mathcal{I}_{x}$ are further broken by $q_{x}\not=0$.
This device is characterized by a bias-direction-selective transverse current carrier, i.e. the forward-bias Cooper-pair current and reverse-bias single-electron current.
This carrier alteration manifests in the differential conductance between opposite biases in Fig. \ref{fig_current}(b).
For positive bias ($eV\sim0.7\Delta$), Cooper pairs dominate the current in all directions, i.e. $G_{y,\theta _{k}>0}(eV)\approx G_{y,\theta _{k}<0}(eV)\approx 2G_{y,\theta _{k}<0} (eV\gg\Delta)$, resulting in $I_y\approx0$.
While negative bias, Cooper-pair transferring and single-electron transmission processes gain a direction-selective contribution to the current, i.e. $G_{y,\theta _{k}<0}(-eV) \sim 2G_{y,\theta _{k}>0}(-eV)$, causing a non-zero $I_y$.

High rectification efficiency is expected due to this bias-direction-selective Cooper pair transverse transport, which manifests in distinct two aspects.
(i) The rectified \textit{conductance} defined as 
\begin{equation}
\PH{\Gamma _{x/y}( eV ) =\frac{G_{x/y}(
+eV ) -G_{x/y}( -eV
) }{G_{x/y}( +eV ) +G_{x/y}(
-eV ) }}\text{.}  \label{eq_gg}
\end{equation}%
is divergent when $\theta _{q}=\pi /2$ [Fig. \ref{fig_current}(d)], indicating that identical currents are driven regardless of the bias polarity [Fig. \ref{fig_current}(c)]. 
(ii) A fully rectified \textit{current}
\begin{equation}
\eta _{x/y}( e V ) =\frac{\delta
I_{x/y}( eV ) }{|I_{x/y}(+
eV ) |+|I_{x/y}( -eV
) |}\text{,}  \label{eq_eta}
\end{equation}%
is expected [Fig. \ref{fig_current}(e)] when $q_{x,y}\neq0$ and $e|V|/\Delta\sim1$, where
\begin{equation}
\delta I_{x/y} =sign[I_{x/y}(+
eV )]|I_{x/y}(+
eV ) -I_{x/y}( -eV
) |\text{.}
\label{eq_dixy}
\end{equation}%
When the biases are around the superconducting gap, the transverse current is negligible due to the compensation Cooper-pair current flowing perpendicular to the junction under one bias and becomes finite stemming from the direction-selective current of Cooper-pairs and single electrons, as schematic in Fig. \ref{fig_diode}.
These two aspects endow the transverse Cooper-pair diode beyond its longitudinal counterpart, whose diode efficiency is inherently constrained, as it necessitates the transfer of carriers (either Cooper pairs or single electrons) upon reversal of the bias.

The symmetry constrain (Table \ref{Table_Sym}) is revealed in Eq. (\ref{eq_dixy}) and Fig. \ref{fig_current}(f). 
The supercurrent component parallel to the junction direction ($q_{x}=\cos \theta _{q}$) breaks both $\mathcal{I}_x$ and $\mathcal{T}$ and subsequently causes the longitudinal Cooper pair effect embodying in $\delta I_{x}\sim \cos \theta _{q}$. 
While the transverse current difference is $\delta I_{y}\sim \sin 2\theta _{q}$ implying that the addition $\mathcal{M}_y$ broken by the supercurrent component perpendicular to the junction direction ($q_{y}=\sin \theta _{q}$).

%%%%%%%%%%%%%%%%%%%%
%%%%% Figure 4 %%%%%
%%%%%%%%%%%%%%%%%%%%

%\begin{figure}[t]
%    \centering \includegraphics[width=0.48\textwidth]{fig4_acdc.pdf}
%    \caption{Schematic transverse AC-DC converter based on transverse Cooper-pair rectifier. The sinusoidal AC input $V(t) =\Delta/e\sin(\omega t)$ with frequency $\omega$ is rectified into a DC $I_y$ then supports the electric load. The rectified output can be manipulated by the built-in supercurrent, by considering Eq. (\ref{eq_iv}) in quasi-equilibrium.}
    \label{fig_acdc}
%\end{figure}

%%%%%%%%%%%%%%%%%%%%%%%%%%%%%%%%%%%%%
%%%%%%%%%%%%%%%%%%%%%%%%%%%%%%%%%%%%%
%%%%%%%%%%%%%%%%%%%%%%%%%%%%%%%%%%%%%
%\textit{\textcolor{blue}{Diode efficiency.-}}

%%%%%%%%%%%%%%%%%%%%%%%%%%%%%%%%%%%%%
%%%%%%%%%%%%%%%%%%%%%%%%%%%%%%%%%%%%%
%%%%%%%%%%%%%%%%%%%%%%%%%%%%%%%%%%%%%
\textit{\textcolor{blue}{Discussions.-}}
\PH{Experimentally, the transverse Cooper-pair rectifier can be realized in a four-terminal N-S junction with two normal metal probes \cite{sm}. For instance, such setup can be constructed in semiconducting \cite{Strambini2022,Araujo2023,Costa2022,Costa2023,Mazur2022,Turini2022,Jeon2022,Margineda2023a,Ciaccia2023,Reinhardt2023} or topological materials like Bi$_2$Te$_3$-NbSe$_2$ hybrid structure \cite{Zhu2021}. 
One of the promising platforms is the InGaAs/InAs with Al as a proximate superconductor, where the effective built-in supercurrent is contorted by an external in-plane magnetic field interacting with spin-orbit coupling \cite{Kjaergaard2016}. 
The mechanism for generating effective built-in supercurrent or Cooper-pair momentum is also compatible with topological surface states with in-plane Zeeman field and proximate superconductivity \cite{Pal2022,Zhu2021}, such as in NiTe$_2$ \cite{Pal2022} and Bi$_2$Te$_3$ \cite{Zhu2021}. }

\PH{As a proof-of-concept, we simulate \cite{sm} the non-reciprocal current and conductance of a four-terminal device via KWANT \cite{Groth2013KwantAS,Anantram1996}.
The simulation results are consistent with the two-terminal model presented above.
The numerical simulation also demonstrates that the transverse rectifications are robust against finite-thickness effects \cite{Smolyaninova2016EnhancedSI,David2023Magnetic}, systematic inhomogeneous and disorders. }

We remark that the transverse current in the Dirac system such as topological surface states may be helpful to distinguish the retro- and specular AR \cite{Beenakker2006,Salehi2023}.
\PH{Finally, transverse Cooper-pair diode operating in a dynamical regime \cite{Hu2007,Misaki2021} remains an open question that can be investigated in future works.}

%In conclusion, we predict Cooper-pair diodes by developing the BTK formalism in characterizing the $I$-$V$ relation of the N-S junction with a built-in supercurrent. 
%Remarkably, a transverse Cooper-pair diode is expected when the built-in supercurrent breaks $\mathcal{I}_x$, $\mathcal{T}$, and $\mathcal{M}_y$ simultaneously. 
%The transverse Cooper-pair diode enables highly efficient direction-selective Cooper pair transport decoupling the path of the input excitation from the output rectified signal. 
%With fully polarized diode efficiency and colossal nonreciprocal conductance rectification, the proposed diode effects may be of potential application in non-reciprocal superconducting electronics.

%\textit{\textcolor[rgb]{0.00,0.07,1.00}{Acknowledgments.-}}
\begin{acknowledgments}
P.-H. Fu thanks Weiping Xu and Alessandro Braggio for inspiring discussions. 
P.-H. F. \& Y. S. A. are supported by the Singapore Ministry of Education (MOE) Academic Research Fund (AcRF) Tier 2 Grant (MOE-T2EP50221-0019). 
C.H. L. is supported by Singapore’s NRF Quantum engineering grant NRF2021-QEP2-02-P09 and Singapore’s MOE Tier-II grant Proposal ID: T2EP50222-0003.
J.-F. L. is supported by the National Natural Science Foundation of China (Grant No. 12174077) and the Joint Fund with Guangzhou Municipality under Grant No. 202201020238.
Y. Xu is supported by the Scientific Research Starting Foundation of Ningbo University of Technology (Grant No. 2022KQ51).
\end{acknowledgments}

%\bibliography{references}

\end{document}

\title{Supplemental Material for "Transverse Cooper-Pair Rectifier"}
\author{Pei-Hao Fu}
\email{peihao\_fu@sutd.edu.sg}
\affiliation{Science, Mathematics and Technology, Singapore University of Technology and Design, Singapore 487372, Singapore}
\affiliation{Department of Physics, National University of Singapore, Singapore 117542}

\author{Yong Xu}
%\email{xuyong_nbut@163.com}
\affiliation{Institute of Materials, Ningbo University of Technology, Ningbo 315016, China}

\author{Jun-Feng Liu}
\email{phjfliu@gzhu.edu.cn}
\affiliation{School of Physics and Materials Science, Guangzhou University, Guangzhou 510006, China}

\author{Ching Hua Lee}
\email{phylch@nus.edu.sg}
\affiliation{Department of Physics, National University of Singapore, Singapore 117542}

\author{Yee Sin Ang}
\email{yeesin\_ang@sutd.edu.sg}
\affiliation{Science, Mathematics and Technology, Singapore University of Technology and Design, Singapore 487372, Singapore}

\maketitle
\tableofcontents
\newpage
%%%%%%%%%%%%%%%%%%%%%%%%%%%%%%%%%%%%%%%%%%%%%%%%%%%%%%%%%%%%%
\section{Scattering coefficients}
\subsection{Quantum tunneling problem}
\begin{figure}[t]
\centering \includegraphics[width=0.8\textwidth]{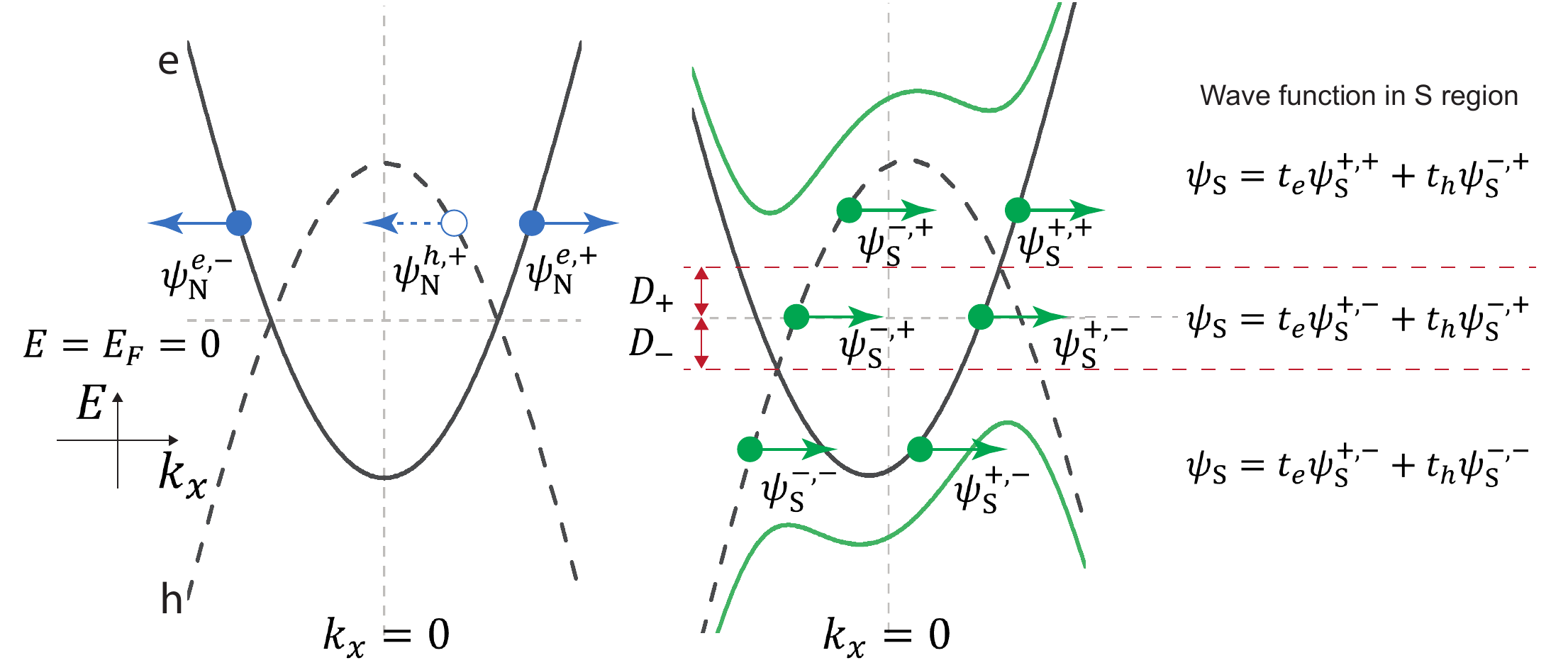}
\caption{Schematic selection rules of the wave function in (left panel) the N site [Eq. (\ref{eq_psiN})] and (right panel) the S site [Eq. (\ref{eq_psiS})] determined by the energy measured from the Doppler energy shift $D_{\pm}$.}
\label{figs_wf}
\end{figure}

The scattering coefficients can be obtained by quantum tunneling problem in the N-S junction with a Hamiltonian    
\begin{eqnarray}
H_{\text{NS}}\left( x\right)  &=&[\hbar ^{2}/(2m)\left( -\partial
_{x}^{2}+k_{y}^{2}\right) -E_{F}+U_{b}\delta \left( x\right) ]\tau _{z}
\label{eq_Hns} \\
&&+\left[ \Delta \tau _{x}+\hbar ^{2}/(2m)\left( q^{2}\tau
_{z}+2q_{y}k_{y}\tau _{0}\right) \right] \Theta \left( x\right)   \nonumber
\\
&&-i\hbar ^{2}/(2m)q_{x}\tau _{0}\left[ \partial _{x}\Theta \left( x\right)
+\Theta \left( x\right) \partial _{x}\right] \text{,}  \nonumber
\end{eqnarray}%
and by integrating the Schr\"odinger equation around the boundary at $x=0$: $%
\int_{0^{-}}^{0^{+}}dxH_{\text{NS}}\left( x\right) \psi \left( x\right)
=\int_{0^{-}}^{0^{+}}dxE\psi \left( x\right) $, the boundary condition is 
\begin{eqnarray}
\psi _{\text{N}}\left( 0\right)  &=&\psi _{\text{S}}\left( 0\right) \text{,}
\label{eq_bdy1} \\
\psi _{\text{S}}^{\prime }\left( 0\right) -\psi _{\text{N}}^{\prime }\left(
0\right)  &=&\left( u_{b}\tau _{0}-iq_{x}\tau _{z}\right) \psi _{\text{N}%
}\left( 0\right) \text{,}  \label{eq_bdy2}
\end{eqnarray}%
where $u_{b}=2mU_{b}/\hbar ^{2}$ and $\psi _{\text{N}}\left( x\right) $ and $%
\psi _{\text{S}}\left( x\right) $ are the wave functions in the N and S side localized in the region of $x<0$ and $x>0$, respectively. 
By solving the wave function of both regions, we obtain 
\begin{eqnarray}
\psi _{\text{N}}\left( x\right)  &=&\psi _{N}^{e,+}+r\psi
_{N}^{e,-}+r_{a}\psi _{N}^{h,+}  \nonumber \\
&=&\left( 
\begin{array}{c}
1 \\ 
0%
\end{array}%
\right) e^{+ik_{+}x}e^{ik_{y}y}+r\left( 
\begin{array}{c}
1 \\ 
0%
\end{array}%
\right) e^{-ik_{+}x}e^{ik_{y}y}+r_{a}\left( 
\begin{array}{c}
0 \\ 
1%
\end{array}%
\right) e^{+ik_{-}x}e^{ik_{y}y}\text{,}  \label{eq_psiN}
\end{eqnarray}%
in the N region where 
\begin{equation}
k_{\pm }=\frac{\sqrt{2m\left( E_{F}\pm E\right) }}{\hbar }\text{,}
\label{eq_kn}
\end{equation}%
is the Fermi momentum of electrons ($+$) and holes ($-$), $r$ and $r_{a}$ are the normal and Andreev reflection amplitudes, respectively. 
The wave function in the S region is
\begin{equation}
\psi _{\text{S}}\left( x\right) =t_{e}\psi _{\text{S}}^{e}+t_{h}\psi _{\text{%
S}}^{h}\text{,}  \label{eq_psiS}
\end{equation}%
where $t_{e}$ and $t_{h}$ are the transmission probability of electron-like quasiparticle $\psi _{\text{S}}^{e}$ and hole-like quasiparticle $\psi _{\text{S}}^{h}$. 
The quasiparticle wave functions are determined by the energy measured from the Doppler energy shift $D_{\pm}$, illustrated in Fig. \ref{figs_wf}, which are 
\begin{equation}
\psi _{\text{S}}^{e}=\Theta \left( E-D_{+}\right) \psi _{S}^{+,+}+\Theta
\left( D_{+}-E\right) \psi _{S}^{+,-}\text{,}  \label{eq_phie}
\end{equation}%
for electron-like quasiparticles and 
\begin{equation}
\psi _{\text{S}}^{h}=\Theta \left( E-D_{-}\right) \psi _{S}^{-,-}+\Theta
\left( D_{-}-E\right) \psi _{S}^{-,+}\text{,}  \label{eq_psih}
\end{equation}%
for hole-like quasiparticles, respectively. 
The components are  
\begin{eqnarray}
\psi _{S}^{\alpha ,\beta } &=&\left( 
\begin{array}{c}
\sqrt{\frac{\epsilon _{\alpha ,\beta }+\beta \Omega _{\alpha ,\beta }}{%
2\epsilon _{\alpha ,\beta }}} \\ 
\sqrt{\frac{\epsilon _{\alpha ,\beta }-\beta \Omega _{\alpha ,\beta }}{%
2\epsilon _{\alpha ,\beta }}}%
\end{array}%
\right) e^{\alpha i\kappa _{\beta }x}e^{ik_{y}y}\text{,}  \label{eq_phiss} \\
\Omega _{\alpha ,\beta } &=&\sqrt{\epsilon _{\alpha ,\beta }^{2}-\Delta ^{2}}%
\text{,}  \label{eq_Ome} \\
\epsilon _{\alpha ,\beta } &=&E-\frac{\hbar ^{2}}{m}q_{y}k_{y}-\frac{\hbar
^{2}}{m}q_{x}\left( \alpha \kappa _{\beta }\right) \text{,}  \label{eq_eps}
\end{eqnarray}%
where the wave vector $\kappa _{\beta }$ in the S region is determined by the equation 
\begin{equation}
\left[ \left( E-\frac{\hbar ^{2}}{m}q_{y}k_{y}\right) -\frac{\hbar ^{2}}{m}%
q_{x}\left( \alpha \kappa _{\beta }\right) \right] ^{2}=\left[ \frac{\hbar
^{2}}{2m}\kappa _{\beta }^{2}-\left( E_{F}-\frac{\hbar ^{2}}{2m}k_{y}^{2}-%
\frac{\hbar ^{2}}{2m}\left( \kappa _{\beta }^{2}+q_{y}^{2}\right) \right) %
\right] ^{2}+\Delta ^{2}\text{.}  \label{eq_ee}
\end{equation}%
%%%%%%%%%%%%%%%%%%%%
%%%%% Figure S1 %%%%%
%%%%%%%%%%%%%%%%%%%%

By substituting Eq. (\ref{eq_psiN}) and (\ref{eq_psiS}) in to the boundary
condiction Eq. (\ref{eq_bdy1}) and (\ref{eq_bdy2}), the normal amplitude $r$%
, Andreev reflection amplitude $r_{a}$, transmission probability of
electron-like quasiparticle $t_{e}$ and hole-like quasiparticle $t_{h}$ can
be obtained. However, the complexity of solving Eq. (\ref{eq_ee}) requires the Andreev approximation \cite{Andreev1964,Blonder1982}.

%%%%%%%%%%%%%%%%%%%%%%%%%%%%%%%%%%%%
%%%%%%%%%%%%%%%%%%%%%%%%%%%%%%%%%%%%
%%%%%%%%%%%%%%%%%%%%%%%%%%%%%%%%%%%%
%%%%%%%%%%%%%%%%%%%%%%%%%%%%%%%%%%%%
\subsection{Andreev approximation}
The Andreev approximation is applied in the strong coupling regime when the Fermi energy to the system is much greater than the superconducting pairing gap and the kinetic energy of the drifting Cooper pairs, i.e. $E_{F}\gg $ $\Delta $, $\hbar ^{2}q^{2}/(2m)$, so that the effect of the spectrum mismatch between N and S region is negligible due to $k_{F}=\sqrt{2mE_{F}}/\hbar \gg q$ and the momentum of all quasiparticles are of identical magnitudes, i.e. 
\begin{eqnarray}
k_{+} &=&k_{-}=\kappa _{\pm }=k_{Fx}=k_{F}\cos \theta _{k}\text{,} \\
k_{y} &=&k_{F_{y}}=k_{F}\sin \theta _{k}\text{,}
\end{eqnarray}%
with $\theta _{k}\in \left[ -\pi /2,\pi /2\right] $. As a result, Eq. (\ref%
{eq_phie}) and (\ref{eq_psih}) are reduced as 
\begin{equation}
\psi _{\text{S}}^{e}=\left( 
\begin{array}{c}
\gamma _{+}^{\text{e}} \\ 
\gamma _{-}^{\text{e}}%
\end{array}%
\right) e^{+ik_{Fx}x}e^{ik_{Fy}y}\text{,}
\end{equation}%
and%
\begin{equation}
\psi _{\text{S}}^{h}=\left( 
\begin{array}{c}
\gamma _{-}^{\text{h}} \\ 
\gamma _{+}^{\text{h}}%
\end{array}%
\right) e^{-ik_{Fx}x}e^{ik_{Fy}y}\text{,}
\end{equation}%
respectively, where 
\begin{eqnarray}
\gamma _{\tau }^{\text{e(h)}} &=&\sqrt{1+(-)\tau \frac{\Omega _{\tau }}{%
\epsilon _{\tau }}}\text{,} \\
\epsilon _{\tau } &=&E-D_{\tau }-\frac{\hbar ^{2}}{2m}q^{2}\text{,} \\
\Omega _{\text{e(h)}} &=&sign(E-D_{\tau })\sqrt{\epsilon _{\tau }^{2}-\Delta
^{2}}\text{.}
\end{eqnarray}

With Andreev approximation, the effect of the built-in supercurrent is
absorbed in the $\gamma _{\tau }^{\text{e(h)}}$ through the Dopper energy
shift $D_{\tau }$. The resutling scattering amplituedes are 
\begin{eqnarray}
r &=&Z^{-1}\sum_{\tau }(-\tau )[V_{0}^{2}+q_{x}^{2}+2k_{Fx}(iu_{b}+\tau
q_{x})]\gamma _{\tau }^{\text{e}}\gamma _{\tau }^{\text{h}}\text{,} \\
r_{a} &=&4Z^{-1}\gamma _{-}^{\text{e}}\gamma _{+}^{\text{h}}k_{Fx}^{2}\text{,%
} \\
t_{e} &=&2Z^{-1}\gamma _{+}^{\text{h}}k_{Fx}(2k_{Fx}+q_{x}-iu_{b})\text{,} \\
t_{h} &=&-2Z^{-1}\gamma _{-}^{\text{e}}k_{F_{x}}(q_{x}-iu_{b})\text{,}
\end{eqnarray}%
with 
\begin{equation}
Z=\gamma _{+}^{\text{e}}\gamma _{+}^{\text{h}%
}[u_{b}^{2}+(2k_{Fx}+q_{x})^{2}]-\gamma _{-}^{\text{e}}\gamma _{-}^{\text{h}%
}(u_{b}^{2}+q_{x}^{2})\text{.}
\end{equation}

Remarkably, This approximation is valid for angles of incidence $\theta
_{k}\lesssim \pi /2$ but becomes inaccurate in the vicinity of $\theta
_{k}\sim \pi /2$. However, quasiparticles with momentum nearly parallel to
the interface do not contribute significantly to the current and their
effect in the $I$-$V$ relation and related quantities will thus be
negligible. We have in fact checked our results numerically using lattice
Green's function method (see below) and found they are qualitatively
consistent with the analytical ones with negligible quantitative differences.

\subsection{Current operators and scattering coefficiencts}

\begin{figure}[t]
\centering \includegraphics[width=0.8\textwidth]{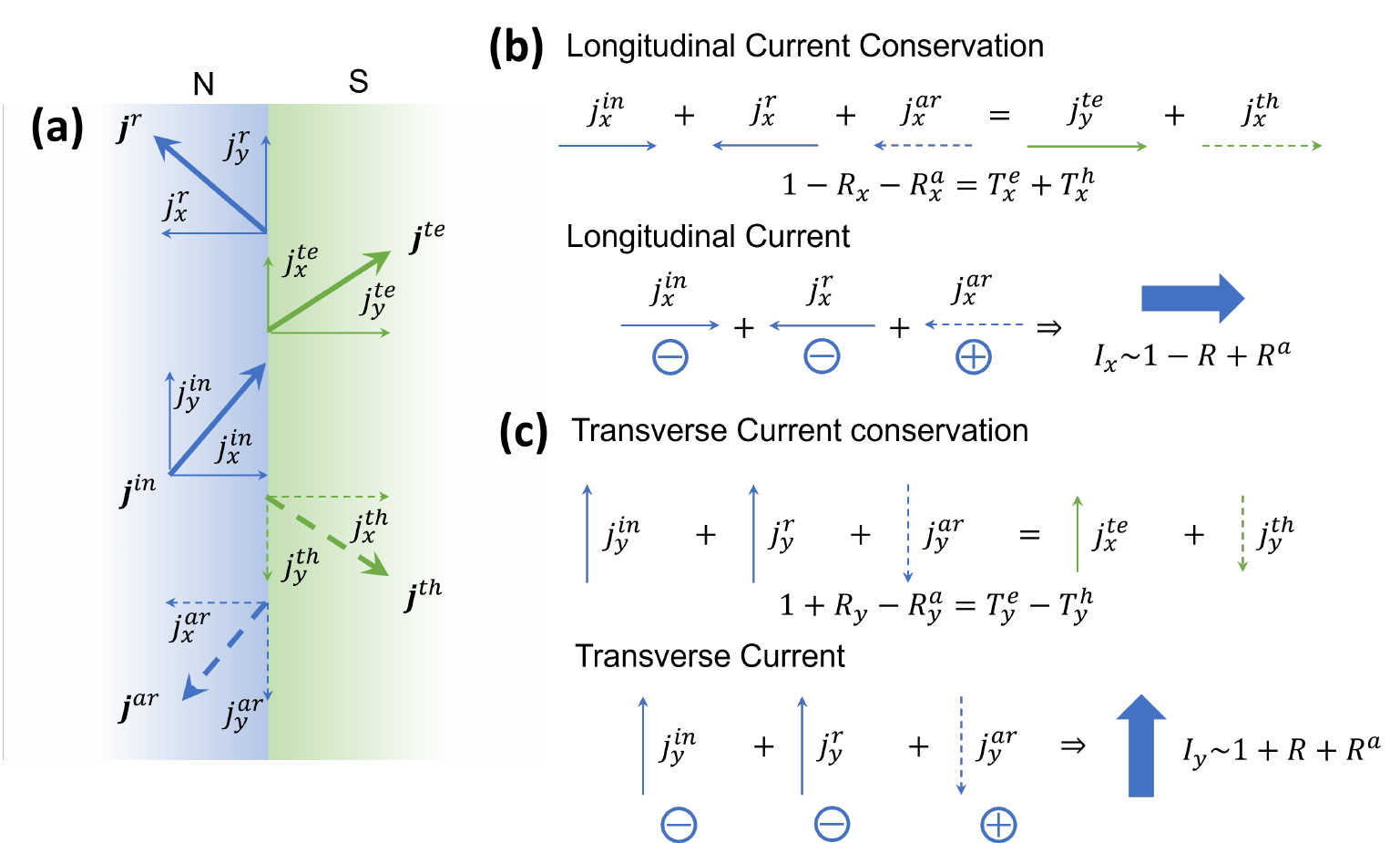}
\caption{(a) Schematic quasiparticle current [Eq. (\ref{eq_jin}-\ref{eq_jth})] in N (blue) and S (green) side and their components in the longitudinal ($x$) and transverse ($y$) directions. 
(b) Schematic longitudinal current conservation [Eq. (\ref{eq_pX})] and quasiparticle current contribution to the longitudinal current [Eq. (\ref{eq_Ix})].
The normal reflection reduces the longitudinal current because of the retro-scattering, while the Andreev reflection contributes to the current by dragging electrons to form Cooper pairs, which manifests in a reflected hole with an opposite charge to the incident electron.
(c) Schematic transverse current conservation [Eq. (\ref{eq_pY})] and quasiparticle current contribution to the transverse current [Eq. (\ref{eq_Iy}). Both the normal specular reflection and the Andreev retro-reflection contribute to the transverse current due to their opposite charges.
}
\label{figs_ScMtI}
\end{figure}

From the Hamiltonian [Eq. (\ref{eq_Hns})], the current operator in the N region is 
\begin{equation}
\bm{j}_{\text{N}}\left( E,\theta _{k}\right) =\frac{\hbar }{m}\text{Im}\left[
\psi _{\text{N}}^{\dagger }\left( x\right) \bm{\nabla }\tau _{z}\psi _{\text{N}}\left(
x\right) \right] \text{,}
\end{equation}%
and in the S region is 
\begin{equation}
\bm{j}_{\text{S}}\left( E,\theta _{k}\right) =\frac{\hbar }{m}\left[ \text{Im}%
\left[ \psi _{\text{S}}\left( x\right) \bm{\nabla }\tau _{z}\psi _{\text{S}}\left(
x\right) \right] +\bm{q}\psi_{\text{S}} ^{\dagger }\left( x\right) \tau _{0}\psi_{\text{S}}
\left( x\right) \right] \text{.}
\end{equation}%
Thus the expectation value of the scattered quasiparticle current can be obtained from the current operator. 
For example, the current of the incident is $\bm{j}%
^{in}=\left( j_{x}^{in},j_{y}^{in}\right) $  
\begin{eqnarray}
j_{x}^{in}\left( E,\theta _{k}\right)  &=&j_{\text{N},x}^{e,+}=\frac{\hbar }{m}%
\text{Im}\left[ \left( \psi _{\text{N}}^{e,+}\right) ^{\dagger }\bm{\nabla }%
\tau _{z}\psi _{\text{N}}^{e,+}\right] =\frac{\hbar }{m}k_{F_{x}}\text{,} \label{eq_jin} \\
j_{y}^{in}\left( E,\theta _{k}\right)  &=&j_{\text{N},y}^{e,+}=\frac{\hbar }{m}%
k_{F_{y}}\text{,} 
\end{eqnarray}%
Similarly, the current of the reflected electron is $\bm{j}^{r}=\left(
j_{x}^{r},j_{y}^{r}\right) $ with 
\begin{equation}
\begin{array}{ccc}
j_{x}^{r}\left( E,\theta _{k}\right) =j_{\text{N},x}^{e,-}=-\frac{\hbar }{m}%
k_{F_{x}}r^{\dagger }r & \text{,} & j_{y}^{r}\left( E,\theta _{k}\right)
=j_{\text{N},y}^{e,-}=\frac{\hbar }{m}k_{F_{y}}r^{\dagger }r%
\end{array}%
\text{,}
\end{equation}%
the current from the Andreev reflected hole is $\bm{j}^{ar}=\left(
j_{x}^{ar},j_{y}^{ar}\right) $ with%
\[
\begin{array}{ccc}
j_{x}^{ar}\left( E,\theta _{k}\right) =j_{\text{N},x}^{h,+}=-\frac{\hbar ^{2}}{m}%
k_{F_{x}}r_{a}^{\dagger }r_{a} & \text{,} & j_{y}^{ar}\left( E,\theta
_{k}\right) =j_{\text{N},y}^{h,+}=-\frac{\hbar ^{2}}{m}k_{F_{y}}r_{a}^{\dagger
}r_{a}%
\end{array}%
\text{,}
\]%
the current from the transmitted electron-like quasiparticle is $\bm{j}%
^{te}=\left( j_{x}^{te},j_{y}^{te}\right) $ with%
\begin{equation}
\begin{array}{ccc}
j_{x}^{te}\left( E,\theta _{k}\right) =\frac{\hbar }{m}t_{e}^{\dagger
}t_{e}\left( q_{x}+\text{Re}\frac{k_{F_{x}}\Omega _{+}}{\epsilon _{+}}%
\right)  & \text{,} & j_{y}^{te}\left( E,\theta _{k}\right) =\frac{\hbar }{m}%
t_{e}^{\dagger }t_{e}\left( q_{y}+\text{Re}\frac{k_{F_{y}}\Omega _{+}}{%
\epsilon _{+}}\right) 
\end{array}%
\text{,}
\end{equation}%
and current from the transimitted hole-like quasiparticle is $\bm{j}%
^{te}=\left( j_{x}^{th},j_{y}^{th}\right) $ with%
\begin{equation}
\begin{array}{ccc}
j_{x}^{th}\left( E,\theta _{k}\right) =\frac{\hbar }{m}t_{h}^{\dagger
}t_{h}\left( q_{x}+\text{Re}\frac{k_{F_{x}}\Omega _{-}}{\epsilon _{-}}%
\right)  & \text{,} & j_{y}^{th}\left( E,\theta _{k}\right) =\frac{\hbar }{m}%
t_{h}^{\dagger }t_{h}\left( q_{y}-\text{Re}\frac{k_{F_{y}}\Omega _{-}}{%
\epsilon _{-}}\right) 
\end{array} \label{eq_jth}
\text{.}
\end{equation}%
All the currents are schematically exhibited in Fig. \ref{figs_ScMtI}. 

The scattering coefficients probability are defined by the ratio between the scattered current to the incident current, which are 
\begin{equation}
\begin{array}{cc}
R_{x}\left( E,\theta _{k}\right) =\left\vert \frac{j_{x}^{r}}{j_{x}^{in}}%
\right\vert =\left\vert r\right\vert ^{2}\text{,} & R_{x}^{a}\left( E,\theta
_{k}\right) =\left\vert \frac{j_{x}^{ar}}{j_{x}^{in}}\right\vert =\left\vert
r_{a}\right\vert ^{2}\text{,} \\ 
T_{x}^{e}\left( E,\theta _{k}\right) =\left\vert \frac{j_{x}^{te}}{j_{x}^{in}%
}\right\vert =t_{e}^{\dagger }t_{e}\left\vert \frac{q_{x}+\text{Re}\frac{%
k_{F_{x}}\Omega _{+}}{\epsilon _{+}}}{k_{F_{x}}}\right\vert \text{,} & 
T_{x}^{h}\left( E,\theta _{k}\right) =\left\vert \frac{j_{x}^{th}}{j_{x}^{in}%
}\right\vert =t_{h}^{\dagger }t_{h}\left\vert \frac{q_{x}+\text{Re}\frac{%
k_{F_{x}}\Omega _{-}}{\epsilon _{-}}}{k_{F_{x}}}\right\vert \text{,}%
\end{array} \label{eq_scX}
\end{equation}%
in the $x$ direction and 
\begin{equation}
\begin{array}{cc}
R_{y}\left( E,\theta _{k}\right) =\left\vert \frac{j_{re}^{y}}{j_{in}^{y}}%
\right\vert =\left\vert r\right\vert ^{2}\text{,} & R_{y}^{a}\left( E,\theta
_{k}\right) =\left\vert \frac{j_{ar}^{y}}{j_{in}^{y}}\right\vert =\left\vert
r_{a}\right\vert ^{2}\text{,} \\ 
T_{y}^{e}\left( E,\theta _{k}\right) =\left\vert \frac{j_{te}^{y}}{j_{in}^{y}%
}\right\vert =t_{e}^{\dagger }t_{e}\left\vert \frac{q_{y}+\text{Re}\frac{%
k_{F_{y}}\Omega _{+}}{\epsilon _{+}}}{k_{F_{y}}}\right\vert \text{,} & 
T_{y}^{h}\left( E,\theta _{k}\right) =\left\vert \frac{j_{th}^{y}}{j_{in}^{y}%
}\right\vert =t_{h}^{\dagger }t_{h}\left\vert \frac{q_{y}-\text{Re}\frac{%
k_{F_{y}}\Omega _{-}}{\epsilon _{-}}}{k_{F_{y}}}\right\vert \text{,}%
\end{array} \label{eq_scY}
\end{equation}

As illustrated in Fig. \ref{figs_ScMtI}, probability conservation is preserved in both directions relating to all the scattering coefficients. 
In $x$-direction, probabilities between the incoming electrons and the scattered quasiparticles satisfy 
\begin{equation}
1-R_{x}\left( E,\theta _{k}\right) -R_{x}^{a}\left( E,\theta _{k}\right)
=T_{x}^{e}\left( E,\theta _{k}\right) +T_{x}^{h}\left( E,\theta _{k}\right) 
\text{.}  \label{eq_pX}
\end{equation}%
Meanwhile, because the translational symmetry in the $y$ direction is preserved, quasiparticle currents on two sides of the N-SC interface are identical, i.e. 
\begin{equation}
1+R_{y}\left( E,\theta _{k}\right) -R_{y}^{a}\left( E,\theta _{k}\right)
=T_{y}^{e}\left( E,\theta _{k}\right) -T_{y}^{h}\left( E,\theta _{k}\right) 
\text{.}  \label{eq_pY}
\end{equation}%
Moreover, $R_{x}\left( E,\theta _{k}\right) =R_{y}\left( E,\theta
_{k}\right) =R\left( E,\theta _{k}\right) $ and $R_{x}^{a}\left( E,\theta
_{k}\right) =R_{y}^{a}\left( E,\theta _{k}\right) =R_{a}\left( E,\theta
_{k}\right) $ are expected because of the isotropic N side.
%%%%%%%%%%%%%%%%%%%%
%%%%% Figure S2 %%%%%
%%%%%%%%%%%%%%%%%%%%

%%%%%%%%%%%%%%%%%%%%
%%%%%%%%%%%%%%%%%%%%
%%%%%%%%%%%%%%%%%%%%
\section{Generalized BTK equation}
In the BTK formalism \cite{Blonder1982}, the current in the N-S junction can be calculated from the net current through the interface on the N side,  
\begin{equation}
I=e\left( J_{\rightarrow }-J_{\leftarrow }\right) \text{,}
\end{equation}%
where $J_{\rightarrow \text{(}\leftarrow \text{)}}$ denotes in-coming (out-going) quasiparticle current. 
Moreover, the electrochemical potential of the pairs in the superconductor is set as a reference level, i.e., $f_{\text{S}}\left( E\right) =f_{0}\left( E\right) =[1+\exp
(E/k_{B}T)]^{-1}$ and thus $f_{\text{N}}\left( E\right) =f_{0}\left(
E-eV\right) $ for electrons and $1-f_{\text{N}}\left( -E\right) =f_{0}\left(
E+eV\right) $ for holes.
In this section, we implement the idea of the BTK formalism to calculate the transverse current in the N-S junction.
%%%%%%%%%%%%%%%%%%%%
%%%%%%%%%%%%%%%%%%%%
%%%%%%%%%%%%%%%%%%%%
\subsection{Longitudinal current}
Let's first review the BTK formalism for the longitudinal current, which is 
\begin{equation}
i_{x}\left( \theta _{k}\right) =ev_{F_{x}}\left[ J_{\rightarrow ,x}\left(
\theta _{k}\right) -J_{\leftarrow ,x}\left( \theta _{k}\right) \right] \text{%
,}  \label{eq_ix}
\end{equation}%
where 
\begin{equation}
J_{\rightarrow ,x}\left( \theta _{k}\right) =\int_{-\infty }^{+\infty
}dEf_{N}\left( E\right) \text{,}
\end{equation}%
and 
\begin{equation}
J_{\leftarrow ,x}\left( \theta _{k}\right) =\int_{-\infty }^{+\infty }dE
\left[ R_{x}\left( E,\theta _{k}\right) f_{\text{N}}\left( E\right)
+R_{x}^{a}\left( E,\theta _{k}\right) \left[ 1-f_{\text{N}}\left( -E\right) %
\right] +\left[ T_{x}^{e}\left( E,\theta _{k}\right) +T_{x}^{h}\left(
E,\theta _{k}\right) \right] f_{\text{S}}\left( E\right) \right] \text{.}
\end{equation}%
With the probability conservations [Eq. (\ref{eq_pX})], Eq. (\ref{eq_ix}) is
written as 
\begin{equation}
i_{x}\left( \theta _{k}\right) =ev_{F_{x}}\int_{-\infty }^{+\infty }dE\left[
1-R_{x}\left( E,\theta _{k}\right) \right] \left[ f_{\text{N}}\left(
E\right) -f_{\text{S}}\left( E\right) \right] -R_{x}^{a}\left( E,\theta
_{k}\right) \left[ 1-f_{\text{N}}\left( -E\right) -f_{\text{S}}\left(
E\right) \right] \text{.}  \label{eq_ix1}
\end{equation}

In the original BTK paper \cite{Blonder1982}, using the property of the Fermi-Dirac distribution function $f_{\text{S}}\left( E\right) =1-f_{\text{S}}\left( -E\right) $ and
the \textit{symmetry} of the AR $R_{x}^{a}\left( E,\theta _{k}\right)
=R_{x}^{a}\left( -E,\theta _{k}\right) $, the BTK formalism is restored.
However, the asymmetry BdG spectrum involved in this work cause an asymmetric $R_{x}^{a}\left( E,\theta _{k}\right) \neq R_{x}^{a}\left( -E,\theta_{k}\right) $, so one may suspect that could be a modification of the BTK formalism with $i_{x}\left( \theta _{k}\right) \sim \left[ 1-R_{x}\left(
E,\theta _{k}\right) +R_{x}^{a}\left( -E,\theta _{k}\right) \right] \left[
f_{\text{N}}\left( E\right) -f_{\text{S}}\left( E\right) \right] $. 

That is not the case! From the argument by S. Datta, et al, $%
-R_{x}^{a}\left( E,\theta _{k}\right) \left[ 1-f_{\text{N}}\left( -E\right)
-f_{\text{S}}\left( E\right) \right] $ is the current contributed due to the transmission from holes to electrons, which is \textit{symmetry} with the one from electrons to holes, i.e.%
\begin{eqnarray}
j_{h\rightarrow e}\left( E\right)  &=&-\int_{-\infty }^{+\infty
}dER_{x}^{a}\left( E,\theta _{k}\right) \left[ 1-f_{\text{N}}\left(
-E\right) -f_{\text{S}}\left( E\right) \right]   \nonumber \\
&=&-\int_{-\infty }^{+\infty }dER_{x}^{a}\left( -E,\theta _{k}\right) \left[
1-f_{\text{N}}\left( -E\right) -1+f_{\text{S}}\left( -E\right) \right]  
\nonumber \\
&=&-\int_{-\infty }^{+\infty }dER_{x}^{a}\left( -E,\theta _{k}\right) \left[
-f_{\text{N}}\left( -E\right) +f_{\text{S}}\left( -E\right) \right]  
\nonumber \\
&=&-\int_{-\infty }^{+\infty }dER_{x}^{a}\left( -E,\theta _{k}\right) \left[
-f_{\text{N}}\left( -E\right) +f_{\text{S}}\left( -E\right) \right]  \\
&=&\int_{-\infty }^{+\infty }dER_{x}^{a}\left( -E,\theta _{k}\right) \left[
f_{\text{N}}\left( -E\right) -f_{\text{S}}\left( -E\right) \right]  
\nonumber \\
&=&\int_{-\infty }^{+\infty }dER_{x}^{a}\left( E,\theta _{k}\right) \left[
f_{\text{N}}\left( E\right) -f_{\text{S}}\left( E\right) \right]   \nonumber
\\
&=&j_{e\rightarrow h}\left( -E\right) \text{.}  \nonumber
\end{eqnarray}%
Thus Eq. (\ref{eq_ix1}) is further reduced as 
\begin{equation}
i_{x}\left( \theta _{k}\right) =ev_{F_{x}}\int_{-\infty }^{+\infty }dE\left[
1-R_{x}\left( E,\theta _{k}\right) +R_{x}^{a}\left( E,\theta _{k}\right) %
\right] \left[ f_{\text{N}}\left( E\right) -f_{\text{S}}\left( E\right) %
\right]   \label{eq_ix2}
\end{equation}%
which is the BTK formalism. 
This argument justifies the BTK formalism is valid in the case with asymmetric AR $R_{x}^{a}\left( E,\theta
_{k}\right) \neq R_{x}^{a}\left( -E,\theta _{k}\right) $ and the present of built-in supercurrent \cite{Riedel1999,Bagwell1994}, although the original argument assumed the AR is symmetric due to the BdG spectrum. 
After considering all transverse modes the current of the N-S junction in $x$-direction is 
\begin{eqnarray}
I_{x} &=&\frac{Wk_{F}}{2\pi }\int_{-\pi /2}^{\pi /2}i_{x}\left( \theta
_{k}\right) \cos \theta _{k}d\theta _{k}  \nonumber \\
&=&I_{0}\int_{-\infty }^{+\infty }d\frac{E}{\Delta }\int_{-\pi /2}^{\pi /2}%
\left[ 1-R_{x}\left( E,\theta _{k}\right) +R_{x}^{a}\left( E,\theta
_{k}\right) \right] \left[ f_{\text{N}}\left( E\right) -f_{\text{S}}\left(
E\right) \right] \cos \theta _{k}\cos \theta _{k}d\theta   \label{eq_Ix}
\end{eqnarray}%
where $I_{0}=$ $e\Delta Wk_{F}v_{F}/\left( 2\pi \right) \ $the current unit
and the energy $E$ is calculated in the unit of $\Delta $, $%
v_{F_{x}}=v_{F}\cos \theta _{k}=\hbar k_{F}/m\cos \theta _{k}$ is
considered. At zero temperature and small bias, Eq. (\ref{eq_Ix}) can be linearized as 
\begin{equation}
I_{x}=G_{x}V\text{,}
\end{equation}%
with a conductance  
\begin{equation}
G_{x}=G_{0}\int_{-\pi /2}^{\pi /2}\left[ 1-R_{x}\left( eV,\theta _{k}\right)
+R_{x}^{a}\left( eV,\theta _{k}\right) \right] \cos \theta _{k}\cos \theta
_{k}d\theta \text{,}
\end{equation}%
in the unit of $G_{0}=eI_{0}$, where the relation 
\begin{equation}
f_{\text{N}}\left( E\right) -f_{\text{S}}\left( E\right) =\Theta \left(
E\right) \Theta \left( eV-E\right) =\left( -\frac{df\left( E\right) }{dE}%
\right) eV=\delta \left( E-eV\right) eV\text{,}
\end{equation}%
is used in small bias. 

%%%%%%%%%%%%%%%%%%%%%%%%%%%%%%%
%%%%%%%%%%%%%%%%%%%%%%%%%%%%%%%
%%%%%%%%%%%%%%%%%%%%%%%%%%%%%%%
%%%%%%%%%%%%%%%%%%%%%%%%%%%%%%%
\subsection{Transverse current}
A similar argument can be implemented to the transverse current
\begin{equation}
i_{y}\left( \theta _{k}\right) =ev_{F_{y}}\left[ j_{\rightarrow ,y}\left( \theta _{k}\right)-j_{\leftarrow
,y}\left( \theta _{k}\right)\right] \text{,}  \label{eq_iy}
\end{equation}%
where 
\begin{equation}
j_{\rightarrow }^{y}\left( \theta _{k}\right) =\int_{-\infty
}^{+\infty }dE\left[ 1+R_{y}\left( E,\theta _{k}\right) \right] f_{\text{N}%
}\left( E\right) \text{,}
\end{equation}%
with $v_{F_{y}}=v_{F}\sin \theta _{k}$ and 
\begin{eqnarray}
j_{\leftarrow }^{y} &=&\int_{-\infty }^{+\infty }dE\left[
R_{y}^{a}\left( E,\theta _{k}\right) \left[ 1-f_{N}\left( -E\right) \right] +%
\left[ T_{y}^{e}\left( E,\theta _{k}\right) -T_{y}^{h}\left( E,\theta
_{k}\right) \right] f_{S}\left( E\right) \right]  \\
&=&\int_{-\infty }^{+\infty }dE\left[ R_{y}^{a}\left( E,\theta
_{k}\right) \left[ 1-f_{N}\left( -E\right) \right] +\left[ 1+R_{y}\left(
E,\theta _{k}\right) -R_{y}^{a}\left( E,\theta _{k}\right) \right]
f_{S}\left( E\right) \right] \text{,}  \nonumber
\end{eqnarray}%
after considering translational symmetry in the $y$ direction [Eq. (\ref%
{eq_pY})]. Thus Eq. (\ref{eq_iy}) is rewritten as  
\begin{equation}
i_{y}\left( \theta _{k}\right) =ev_{F_{y}}\int_{-\infty }^{+\infty }dE\left[
1+R_{y}\left( E,\theta _{k}\right) +R_{y}^{a}\left( E,\theta _{k}\right) %
\right] \left[ f_{\text{N}}\left( E\right) -f_{\text{S}}\left( E\right) %
\right] \text{,}
\end{equation}%
and the current of the N-S junction in $y$-direction is 
\begin{equation}
I_{y}=I_{0}\int_{-\infty }^{+\infty }d\frac{E}{\Delta }\int_{-\pi /2}^{\pi
/2}\left[ 1+R_{y}\left( E,\theta _{k}\right) +R_{y}^{a}\left( E,\theta
_{k}\right) \right] \left[ f_{\text{N}}\left( E\right) -f_{\text{S}}\left(
E\right) \right] \cos \theta _{k}\sin \theta _{k}d\theta _{k}\text{.}
\label{eq_Iy}
\end{equation}%
The transverse conductance can be obtained by linearizing Eq. (\ref{eq_Iy}),
which is  
\begin{equation}
G_{y}=G_{0}\int_{-\pi /2}^{\pi /2}\left[ 1+R_{y}\left( eV,\theta _{k}\right)
+R_{y}^{a}\left( eV,\theta _{k}\right) \right] \cos \theta _{k}\sin \theta
_{k}d\theta _{k}\text{,}
\end{equation}%
and $I_{y}=G_{y}V$. 

Therefore, Eq. (\ref{eq_Ix}) and (\ref{eq_Iy}) can be written in one
equation as 
\begin{equation}
\bm{I}=I_{0}\int_{-\infty }^{+\infty }d\frac{E}{\Delta }\int_{-\pi
/2}^{\pi /2}\bm{\hat{k}}_{F}\cdot \bm{T}\left( E,\theta _{k}\right) %
\left[ f_{\text{N}}\left( E\right) -f_{\text{S}}\left( E\right) \right] \cos
\theta _{k}d\theta _{k}\text{,}
\end{equation}%
where $\bm{\hat{k}}_{F}=\bm{k}_{F}/v_{F}$, $\bm{T}\left(
E,\theta _{k}\right) =\left( T_{x}\left( E,\theta _{k}\right) ,T_{y}\left(
E,\theta _{k}\right) \right) $ and  
\begin{eqnarray}
T_{x}\left( E,\theta _{k}\right)  &=&1-R\left( E,\theta _{k}\right)
+R^{a}\left( E,\theta _{k}\right) \text{,} \\
T_{y}\left( E,\theta _{k}\right)  &=&1+R\left( E,\theta _{k}\right)
+R^{a}\left( E,\theta _{k}\right) \text{,}
\end{eqnarray}%
and 
\begin{equation}
\bm{G}=G_{0}\int_{-\pi /2}^{\pi /2}\bm{\hat{k}}_{F}\cdot \bm{T}%
\left( eV, \theta _{k}\right) \cos \theta _{k}d\theta _{k}\text{,}
\end{equation}%
with the linearized current 
\begin{equation}
\bm{I=G}V\text{,}
\end{equation}%
which is Eq. (1) in the main text.

As illustrated in Fig. \ref{figs_ScMtI}, Eq. (\ref{eq_Ix}) and (\ref{eq_Iy}) indicate that the normal reflection reduces the longitudinal current because of the retro-scattering, while the Andreev reflection contributes to the current by dragging electrons to form Cooper pairs, which manifests in a reflected hole with an opposite charge to the incident electron.
On the other hand, both the normal specular reflection and the Andreev retro-reflection contribute to the transverse current due to their opposite charges.
%%%%%%%%%%%%%%%%%%%%%%%%%%%%%%%%%%%%%%%%%%
%%%%%%%%%%%%%%%%%%%%%%%%%%%%%%%%%%%%%%%%%%
%%%%%%%%%%%%%%%%%%%%%%%%%%%%%%%%%%%%%%%%%%
\section{Numerical simulations}
\subsection{Model and formalism}
\begin{figure}[t]
\centering \includegraphics[width=0.8\textwidth]{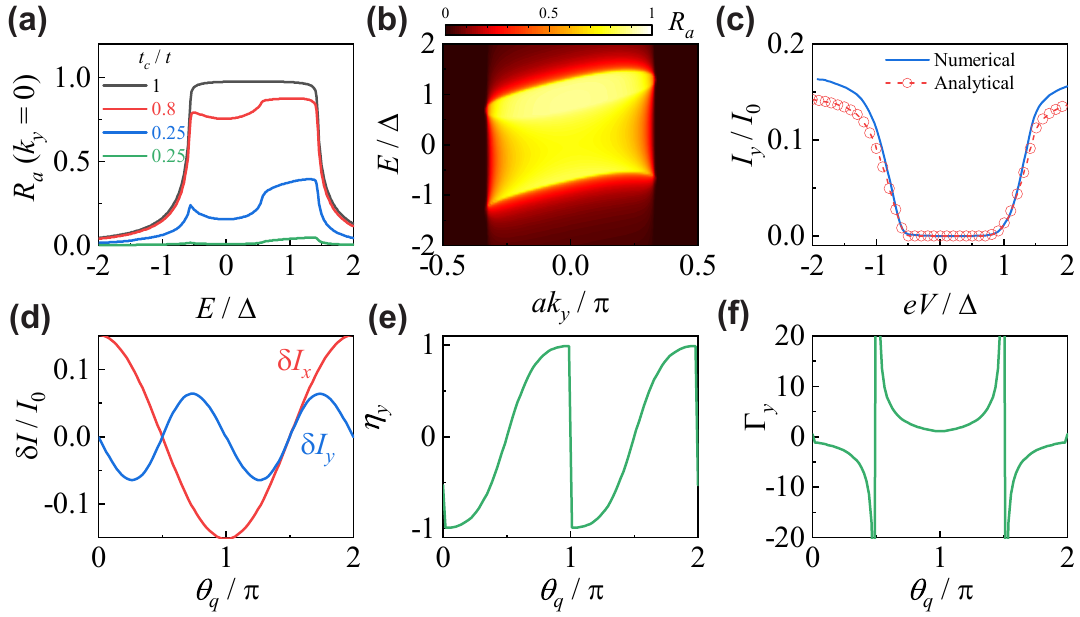}
\caption{Numerical reproduction of the main results demonstrated in the main text, including (a, b) the asymmetric and anisotropic AR, (c) transverse current, (d) current certification with $\delta I_x \sim \cos{\theta_q}$ and $\delta I_y \sim \sin{2\theta_q}$, (e) fully polarized transverse current rectification and (d) colossal nonreciprocal conductance rectification.}
\label{figs_NvsA}
\end{figure}

The analytical results demonstrated in the main text are justified by the numerical one obtained through lattice Green's function technique \cite{Fu2020}, which are exhibited in Fig. \ref{figs_NvsA}.

\begin{figure}[t]
\centering \includegraphics[width=0.8\textwidth]{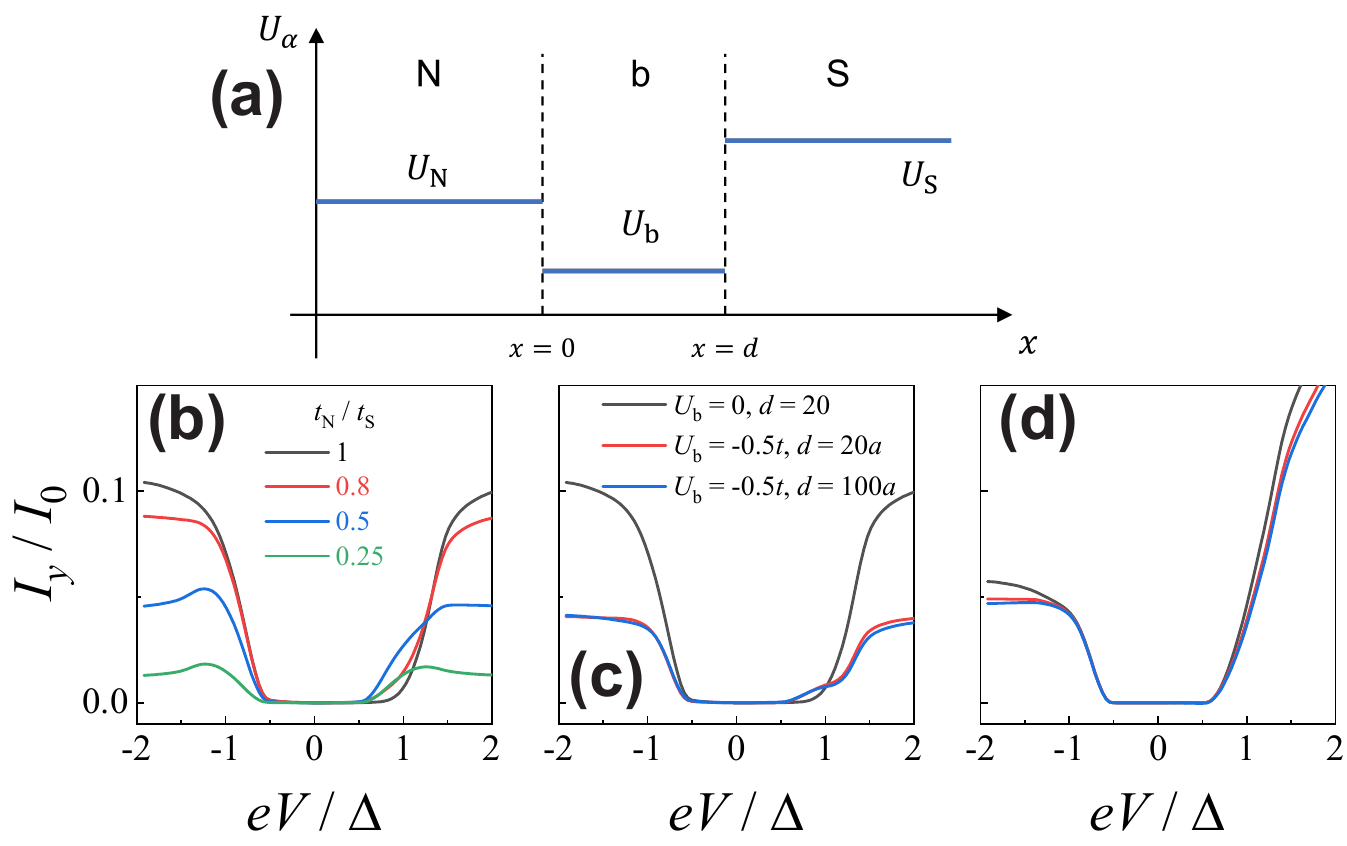}
\caption{(a) Schematic electrical potential profile in N, barrier, and superconducting region, denoted as $U_{\text{N}}$, $U_{\text{b}}$ and $U_{\text{S}}$, respectively. (b-d) Numerical results beyond the assumption (ii-iv). The transverse current is preserved but suppressed when (b) the effective mass $m$ and the Fermi level $E_{F}$ are mismatched between the N and S sides, (c) a barrier separating the N and S region and (d) Andreev approximation is invalid as $E_{F}\gtrsim $ $\Delta $, $\hbar ^{2}q^{2}/(2m)$ with $E_{F}=U_{\text{S}}$ and $\Delta =0.5U_{\text{S}}$.  
}
\label{figs_beyondapp}
\end{figure}

The Hamiltonian of the N-S junction is 
\begin{equation}
H_{\text{latt}}=H_{\text{N}}+H_{\text{b}}+H_{\text{S}}+T_{\text{bS}}+T_{%
\text{Nb}}\text{,}
\end{equation}%
where $H_{\text{N}}$, $H_{\text{b}}$ and $H_{\text{S}}$ are the Hamiltonian
of the normal-metal, barrier, and supercondcting region, respectively, and  
\begin{equation}
H_{\alpha =\text{N,b,S}}=\sum_{\bm{r}}C_{\bm{r}}^{\dagger
}h_{0,\alpha }C_{\bm{r}}+\sum_{\bm{r,\hat{r}}}C_{\bm{r+\hat{r}}%
}^{\dagger }h_{\bm{r},\alpha }C_{\bm{r}}+h.c.\text{,}
\end{equation}%
where $C_{\bm{r}}=\left( c_{\bm{r}},c_{\bm{r}}^{\dagger }\right) 
$ with $c_{\bm{r}}^{\dagger }$\ the creation operator at position $%
\bm{r=}\left( x,y,z\right) $, and $\bm{\hat{r}=}\left( \hat{x},\hat{y%
},\hat{z}\right) $. The components of the Hamiltonian are 
\begin{eqnarray}
h_{0,\alpha } &=&\left( 6t+ta^{2}q^{2}\delta _{\alpha \text{,S}}-U_{\alpha
}\right) \tau _{z}+\Delta \delta _{\alpha \text{,S}}\tau _{x}\text{,} \\
h_{x,\alpha } &=&-t_{\alpha }\tau _{z}-itaq_{x}\delta _{\alpha \text{,S}%
}\tau _{0}\text{,} \\
h_{y,\alpha } &=&-t_{\alpha }\tau _{z}-itaq_{y}\delta _{\alpha \text{,S}%
}\tau _{0}\text{,} \\
h_{z,\alpha } &=&-t_{\alpha }\tau _{z}\text{.}
\end{eqnarray}
The component $T_{\text{bS}}$ ($T_{\text{Nb}}$) descibe the coupling between
the barrier and the supercondctor (normal metal and the barrier), which are  
\begin{equation}
T_{\text{bS}}=\sum_{\bm{r}}C_{\bm{r+\hat{x}}}^{\dagger }(-t_{c}\tau
_{z})C_{\bm{r}}+h.c.\text{,}
\end{equation}%
with a strength $t_{c}$,  $T_{\text{Nb}}=\sum_{\bm{r}}C_{\bm{r+\hat{x%
}}}^{\dagger }(-t_{\text{N}}\tau _{z})C_{\bm{r}}+h.c.$. The potential
profile in the normal, barrier, and superconducting region with $U_{\text{N}}
$, $U_{\text{b}}$ and $U_{\text{S}}$ is shown in Fig. \ref{figs_beyondapp}(a).
\PH{In the barrier region, the potential $U_{\text{b}}$ can be controlled by the tunneling gate \cite{Mazur2022} and the effect of nonmagnetic potential disorder is further included, using the Anderson disorder model $H_{\text{dis}}=\sum_{\bm{r}}C_{\bm{r}}^{\dagger} h_{\text{dis}}\tau_z C_{\bm{r}}$ where the uncorrelated potentials $h_{\text{dis}}$ are drawn from the uniform box distribution on the interval $[-V_{\text{dis}},V_{\text{dis}}]$ and added to every site of our discrete lattice.}

%%%%%%%%%%%%%%%%%%%%%
%%%%% Figure S3 %%%%%
%%%%%%%%%%%%%%%%%%%%%

Using the Fisher-Lee relation, based on the lattice Green's function
technique, the scattering matrix of a junction can be obtained by \cite{Fu2020},
\begin{equation}
S_{q,p}^{\alpha ,\beta }(k_{y})=-\delta _{p,q}\delta _{\alpha ,\beta
}+i[\Gamma _{q}^{\alpha }(k_{y})]^{1/2}G_{q,p}^{\alpha ,\beta
}(k_{y})[\Gamma _{p}^{\beta }(k_{y})]^{1/2}\text{.}
\end{equation}%
Here, $\alpha ,\beta $ $\in \left\{ e,h\right\} $ label the electron or hole and $p,q$ $\in \left\{ \text{N},\text{S}\right\} $ represent the N and S leads. 
The matrix $S_{q,p}^{\alpha,\beta }$ denotes the scattering amplitude of the process where an incident particle $\beta $ from lead $p$ is scattered into a particle $\alpha $ in lead $q$. $\Gamma _{p}^{\beta}(k_{y})$ is a block of the linewidth function $\Gamma _{p}(k_{y}=i[\Sigma_{p}^{r}(k_{y})-\Sigma _{p}^{a}(k_{y})]$ with retarded/advanced self-energy $\Sigma _{p}^{r/a}(k_{y})$ due to the coupling to the lead $p$ which can be calculated numerically by the recursive method. $G_{q,p}^{\alpha,\beta}(k_{y})$ is the matrix block of the retarded Green's function determined by  $G^{r}(k_{y})=[E-h_{\text{b}}(k_{y})-\Sigma _{p}^{r}(k_{y})-\Sigma_{q}^{r}(k_{y})]^{-1}$.
The Andreev reflection coefficient is
\begin{equation}
R_{a}(k_{y})=\text{Tr}[S_{\text{N,N}}^{h,e}(k_{y})]^{\dagger }S_{\text{N,N}%
}^{h,e}(k_{y})\text{.}
\end{equation}%
The other scattering coefficients, $T_{he}(k_{y})$, $T_{ee}(k_{y})$ and $%
R_{ee}(k_{y})$, can also be obtained similarly. 
The current of the N-S junction can be calculated by%
\begin{equation}
I_{x\left( y\right) }=\frac{e^{2}}{h}\sum_{k_{y}}\int_{-\infty }^{+\infty
}dET_{x\left( y\right) }(E,k_{y})\left[ f_{\text{N}}\left( E\right) -f_{%
\text{S}}\left( E\right) \right] \text{,}  \label{g1}
\end{equation}%
with 
\begin{eqnarray}
T_{x}(E,k_{y}) &=&sign\left( \cos ak_{y}\right) \left[
1-R(E,k_{y})-R_{a}(E,k_{y})\right] \text{,} \\
T_{y}(E,k_{y}) &=&sign\left( \sin ak_{y}\right) \left[
1+R(E,k_{y})-R_{a}(E,k_{y})\right] \text{.}
\end{eqnarray}
%%%%%%%%%%%%%%%%%%%%%
%%%%% Figure S4 %%%%%
%%%%%%%%%%%%%%%%%%%%%

\subsection{Beyond Andreev approximation}
The main results in the main text are based on the following four assumptions. 
(i) Translational symmetry in $y$ direction is preserved. 
(ii) The effective mass $m$ and the Fermi level $E_{F}$ are shared by both the N and S sides. 
(iii) The scattering is determined by the interface tunnel barrier $U_{b}$ at $x=0$ which is the only boundary condition involved. 
(iv) Andreev approximation is also applied through $E_{F}\gg $ $\Delta $, $\hbar ^{2}q^{2}/(2m)$. 
Thus, the Fermi wave number mismatch between N and SC region is negligible due to $k_{F}\gg q$.

\PH{In this section, this case beyond the validity of the assumption (ii-iv) is considered in a clean two-dimensional setup, while the effect of finite thickness and disorder is considered in Sec. \ref{sec_finitethickness}. Moreover, the situation beyond assumption (i) may require a four-terminal setup, which is shown in Sec. \ref{sec_exp}. }

The numerical results exhibited in Fig. \ref{figs_NvsA} demonstrate the validity of the assumption (ii-iv) in the analytical ones.
The effective mass $m$ and the Fermi level $E_{F}$ are shared by both the N, barrier and S sides, i.e. $U_{\text{S}}=U_{\text{N}}=U_{\text{b}}$, $t_{\text{S}}=t_{\text{N}}$ to meet assumption (ii) and (iii). the Fermi energy in the S side is fix $E_{F}=U_{\text{S}}=t_{\text{S}}=1$ as the energy unit, \PH{the SC gap is $\Delta =0.001U_{\text{S}}$}, and the Cooper-pair momentum is $q=0.5q_{c}$ with $q_{c}=\Delta/2\hbar $ ($\hbar \equiv 1$) to meet assumption (iv).

The results of the transverse rectifier effect and the performance are robust even when the parameters are chosen beyond the assumptions as shown in Fig. \ref{figs_beyondapp}. 
A transverse rectifier effect is expected when assumption (ii) is invalid because of $t_{\text{S}}\neq t_{\text{N}}$ and/or $U_{\text{S}}\neq U_{\text{N}}=U_{\text{b}%
}$, although the current is suppressed due to the normal reflections enhanced by the Fermi surface mismatches between the N and S sides. 
Moreover, when there is a barrier region with a length $d$ and $U_{\text{b}}\neq U_{\text{S}}=U_{\text{N}}$, assumption (iii) is broken since $x=0$ and $x=d$ are both the boundary in the setup. 
Like the effect of $t_{\text{S}}\neq t_{\text{N}}$ and $U_{\text{S}}\neq U_{\text{N}}=U_{\text{b}}$, the barrier suppresses the current and reduces the asymmetry, but the rectified conductance increases.
Finally, the transverse current certification also exists even when the Andreev approximation is invalid when $E_{F}\gtrsim $ $\Delta $, $\hbar ^{2}q^{2}/(2m)$ with $E_{F}=U_{\text{S}}$ and $\Delta =0.5U_{\text{S}}$.

\subsection{Effect of finite thickness and disorders} \label{sec_finitethickness}

\PH{For finite thickness systems, some characteristic lengths are involved here: the thickness $L_{z}$, the superconducting coherent length $\xi _{s}=\Delta /\hbar v_{F}$ and the London penetration depth $\lambda _{L}$. 
Take aluminum as an example. 
One may obtain $\xi _{s}=1600$ nm \cite{Smolyaninova2016EnhancedSI} and the order of $\lambda _{L}$ ranging from $150$ nm to $50$ nm depending on sample thickness \cite{David2023Magnetic}. 
We argue that the superconductivity and the supercurrent due to the Meissner effect in the N-S junction involved in the main text are inherited from the proximity effect of the aluminum and thus support the same superconducting coherent length and the London penetration depth. }

\PH{The transport properties of the N-S junction are affected by the competition between the sample thickness $L_{z}$ and the London penetration depth $\lambda _{L}$. 
A finite-thickness sample can be numerically considered as a stacking of the two-dimensional systems in the $z$ direction with nearest-neighbor coupling. 
$L_{z}$ denotes the number of layers, each of
which is labeled by $z=1$, $2$, $3$, etc. Since $\xi _{s}\gg \lambda _{L}$, $L_{z}$, the superconducting order parameter can be considered to be layer-independent, $\Delta \left( z\right) =\Delta $, while magnitude of the supercurrent decays as $q_{0z}\left( z\right) =q_{0}\exp \left( -z/\lambda
_{L}\right) $. }

\begin{figure}[t]
\centering \includegraphics[width=0.8\textwidth]{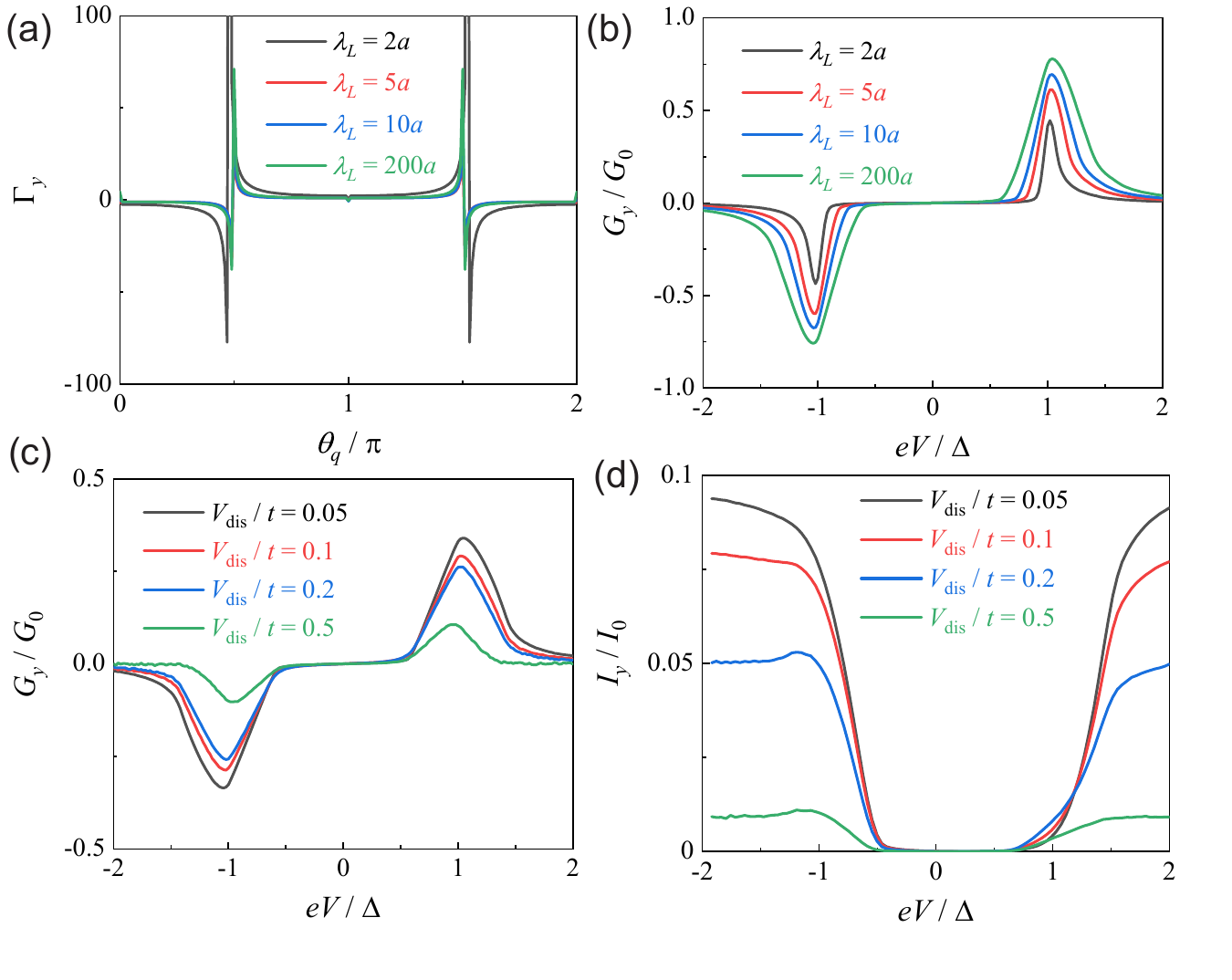}
\caption{\PH{The performance of the transverse Cooper-pair diode in a finite-thickness disorder two-terminal setup, including (a) the conductance rectification efficiency (b, c) conductance, and (d) current. 
The thickness of the sample is set as $L_z=20a$ with $a=1$ as the unit lattice constant. 
In (a) and (b), length of the junction is $L_x=10a$, the London penetration depth is chosen as $\lambda_L=2a\ll L_z$, $\lambda_L=5a ,10a \simeq L_z$ and $\lambda_L=200a\gg L_z$, with the disorder strength $V_{\text{dis}}=0.1t$.
In (c) and (d), the length of the junction is $L_x=100a$, and the London penetration depth is fixed as $\lambda_L=200a$. 
The effect of disorder is averaged after 200 times calculations.
The parameters for effective supercurrent are $q_0=0.5$ and (b, c) $\theta_q=\pi/2$ and (d) $\theta_q=\pi/4$.}
}
\label{figR4_finiteZ}
\end{figure}

\PH{Numerical results in Fig. \ref{figR4_finiteZ} demonstrate the cases with $L_{z}\gg \lambda _{L}$, $L_{z}\simeq \lambda _{L}$ and $L_{z}\ll \lambda _{L}$ with the effect of disorders. 
The colossal nonreciprocal conductance rectification [Fig. \ref{figR4_finiteZ}(a)] is insensitive to $\lambda _{L}$ which suppresses the magnitude of the transverse conductance and the bias region of finite conductance [Fig. \ref{figR4_finiteZ} (b)]. 
The suppression is due to the vanishing contribution from layers with negligible $q_{0z}\left( z\right) $. 
Similar results can be observed with disorders, which enhance the scattering and suppress the transverse conductance [Fig. \ref{figR4_finiteZ}(c)] and the current [Fig. \ref{figR4_finiteZ}(d)].}

\PH{The discussion above is related to the systems with supercurrent induced by the Meissner effect. 
For other experimentally related systems, where, for example the effective supercurrent induced by the interplay between Zeeman field and spin-orbit coupling, similar arguments are applicable for inhomogeneous layer-dependent effect supercurrent distribution.}
%%%%%%%%%%%%%%%%%%%%
%%%%% Figure R4 %%%%%
%%%%%%%%%%%%%%%%%%%%

\subsection{Possible experimental setup}\label{sec_exp}

\begin{figure}[t]
\centering \includegraphics[width=0.8\textwidth]{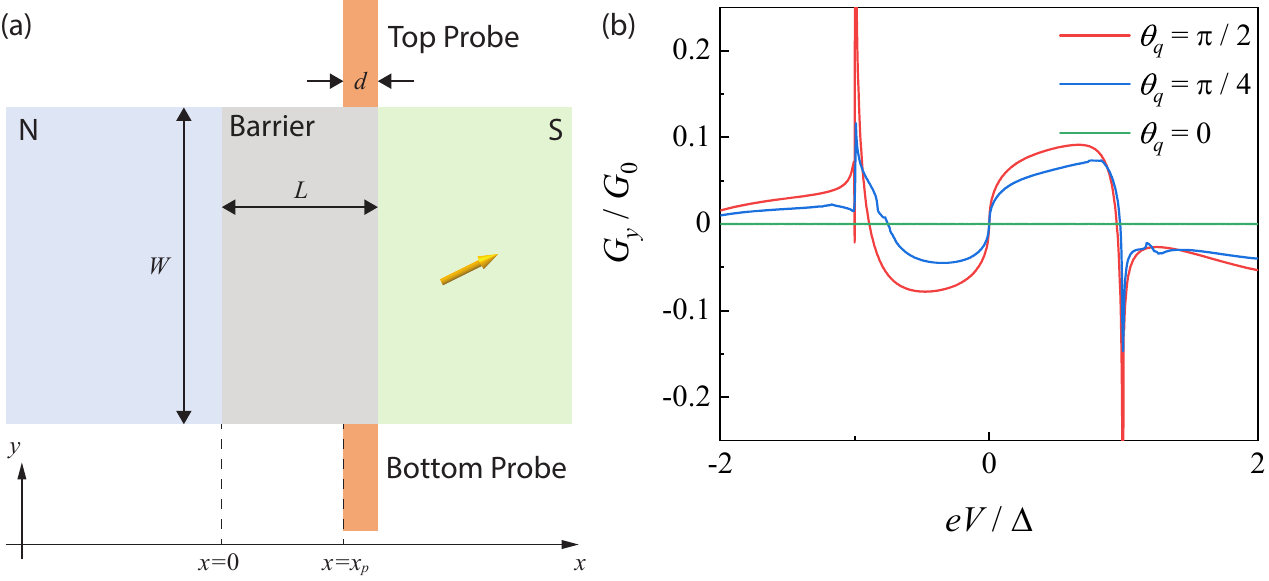}
\caption{\PH{Numerical device simulations using KWANT Package \cite{Groth2013KwantAS}. 
(a) Schematic four-terminal setup for detecting transverse Cooper-pair diode effect. 
The net transverse current of the junction is defined by the current difference between the top and the bottom probes, i.e. $I_{y}=I_{\text{top}}-I_{\text{bottom}}$. 
(b) The conductance [Eq. (\ref{eqr_gy})] to the voltage drop across the N-S junction with various effective supercurrent directions $\theta_q$s. 
The dimensions of the junction are $W=L=20a$, $x_p=18a$, and $d=2a$. }
}
\label{figR3_4tmn}
\end{figure}

\PH{The expected experimental setup for observing the transverse Cooper-pair diode effect is shown in Fig. \ref{figR3_4tmn}. 
The setup is based on a current-biased normal-superconductor (N-S) junction with a voltage drop across this junction, which has been used for the measurement of tunneling spectroscopy in the N-S junction \cite{Strambini2022,Araujo2023}.}

\PH{The proposed transverse Cooper-pair diode effect is expected in this setup after two modifications.}

\PH{(i) A supercurrent is required on the S side of the junction.
The supercurrent can be induced by an additional current injecting into the S side or by the Meissner effect via an in-plane magnetic field \cite{Anthore2003Density}. 
In recent experiments, the supercurrent of the finite-momentum Cooper pairs can be effectively induced by, for example, the interplay between spin-orbit coupling and in-plane Zeeman field and electric field in asymmetric helical states.}

\PH{(ii) Two thin normal metal probes are attached near the N-S interface (at the N side), to observe the transverse Cooper-pair diode effect, as shown in Fig. \ref{figR3_4tmn}(a). 
The current difference between the top and the bottom probes is the net transverse current of the junction, $I_{y}=I_{\text{top}}-I_{\text{bottom}}$.}

\PH{To justify the possibility of the transverse Cooper-pair diode effect, we simulate this two-dimensional four-terminal setup using the KWANT package \cite{Groth2013KwantAS}. 
The Hamiltonian of the system is the same as the two-terminal setup described in the supplemental materials with three normal metal leads and one SC lead. 
The current flowing from the $i^{\text{th}}$ lead is \cite{Anantram1996} 
\begin{equation}
I_{i}=\sum_{j\in N}g_{i,j}\left( \mu _{j}-\mu _{S}\right) \text{,}
\end{equation}%
in the linearized regime (for convenience) with 
\begin{equation}
g_{i,j}=\frac{2e^{2}}{h}\left( \delta
_{i,j}-T_{i,j}^{e,e}+T_{i,j}^{h,e}\right) \text{,}
\end{equation}%
at zero-temperature. The index $i=1$, $2$, $3$, and $4$ denote the left, top, bottom, and right leads, respectively. The left, top, and bottom leads are normal metal and the right lead is the superconductor. 
$\mu _{S}$ and $\mu _{j}$ is the chemical potential in the SC and the $j^{\text{th}}$ lead. 
$T_{i,j}^{\alpha ,\beta }$ with $\alpha ,\beta \in \left\{ e,h\right\} $ is the transmission coefficient of a $\beta $-type quasiparticle in lead $j$ to a $\alpha $-type quasiparticle in lead $i$. With $\mu _{S}=\mu _{2,3}=0$
and $\mu _{1}=eV$, the transverse conductance is 
\begin{equation}
G_{y}=\frac{dI_{y}}{dV}=\frac{d( I_{2}-I_{3}) }{dV}=(
T_{3,1}^{e,e}-T_{2,1}^{e,e}) +(
T_{3,1}^{h,e}-T_{2,1}^{h,e}) \text{,}
\label{eqr_gy}
\end{equation}%
where the positive sign in the second term indicates the hole-like quasiparticle dominates the negative-bias current.}

\PH{The behavior of the transverse conductance shown in Fig. \ref{figR3_4tmn}(b) is basically consistent with the key results obtained in the two-terminal setup demonstrated in the manuscript.
For example, the transverse conductance is of the same magnitude but sign reversal with opposite direction bias when $\theta _{q}=\pi /2$ indicating colossal nonreciprocal conductance rectification.}

\PH{More detailed calculations demonstrate that the transverse conductance rectification efficiency is nearly independent of the width ($W$) and the length ($L$) of the scattering regime but sensitive to the width ($d$) and relative position of the probes as well as their distance ($L-x_p$) to the N-S interface.}
%%%%%%%%%%%%%%%%%%%%
%%%%% Figure R3 %%%%%
%%%%%%%%%%%%%%%%%%%%
%\bibliography{smref.bib}
